\documentclass[twocolumn,tighten]{aastex62}

\usepackage{gensymb}
\usepackage{amsmath}
\usepackage{natbib}

\def\be{\begin{equation}}
\def\en{\end{equation}}

\def\msun{\rm{M_{\odot}}}

\def\mdot{\dot{M}}
\def\H2O{H$_2$O}
\def\CO2{CO$_2$}
\def\DCE45{DCE$_{J-[4.5]}$}
 
\def\Av{$A_{\text{V}} \, $}

\received{---- --, ----}
\revised{---- --, ----}
\accepted{---- --, ----}
\submitjournal{ApJ}

\shorttitle{Inner disk evolution}

\shortauthors{Manzo-Mart{\'i}nez et al.}

\begin{document}

\title{THE EVOLUTION OF THE INNER REGIONS OF PROTOPLANETARY DISKS}

\author[0000-0001-6647-862X]{Ezequiel Manzo-Mart\'inez}
\affiliation{Instituto de Radioastronom\'ia y Astrof\'isica, Universidad Nacional Aut\'onoma de M\'exico, Apartado Postal 3-72, C.P. 58089 Morelia, Michoac\'an, M\'exico}
\author[0000-0002-3950-5386]{Nuria Calvet}
\affiliation{Department of Astronomy, University of Michigan, 323 West Hall, 1085 South University Avenue, Ann Arbor, MI 48109, USA}
\author[0000-0001-9797-5661]{Jes\'us Hern\'andez}
\affiliation{Instituto de Astronom\'ia, Universidad Nacional Aut\'onoma de M\'exico, Ensenada, M\'exico}
\author[0000-0002-2260-7677]{Susana Lizano}
\affiliation{Instituto de Radioastronom\'ia y Astrof\'isica, Universidad Nacional Aut\'onoma de M\'exico, Apartado Postal 3-72, C.P. 58089 Morelia, Michoac\'an, M\'exico}
\author[0000-0002-1650-3740]{Ramiro Franco Hern\'andez}
\affiliation{Instituto de Astronom\'ia y Meteorolog\'ia, Universidad de Guadalajara, M\'exico}
\author{Christopher J. Miller}
\affiliation{Department of Astronomy, University of Michigan, 323 West Hall, 1085 South University Avenue, Ann Arbor, MI 48109, USA}
\author[0000-0001-8284-4343]{Karina Mauc\'o}
\affiliation{N\'ucleo Milenio Formaci\'on Planetaria - NPF, Universidad de Valpara\'iso, Av. Gran Breta\~na 1111, Valpara\'iso, Chile}
\affiliation{Instituto  de  F\'isica  y  Astronom\'ia,  Facultad  de  Ciencias,  Universidad de Valpara\'iso, Av. Gran Breta\~na 1111, 5030 Casilla, Valpara\'iso, Chile}
\author[0000-0001-7124-4094]{C\'esar Brice\~no}
\affiliation{Cerro Tololo Inter-American Observatory, Casilla 603, La Serena 1700000, Chile}
\author{Paola D'Alessio}
\affiliation{Instituto de Radioastronom\'ia y Astrof\'isica, Universidad Nacional Aut\'onoma de M\'exico, Apartado Postal 3-72, C.P. 58089 Morelia, Michoac\'an, M\'exico}

\begin{abstract}
We present a study of the evolution of the inner few au of protoplanetary disks around low-mass stars. We consider nearby stellar groups with ages spanning
from 1 to 11 Myr, distributed into four age bins.
 Combining PANSTARSS photometry with spectral types we derive the reddening consistently for each star, which we use  (1) to measure the excess emission above the photosphere with a new indicator of IR excess and (2) to estimate the mass accretion rate ($\dot{M}$) from the equivalent width of the {H$\alpha$} line.
Using the observed decay of $\dot{M}$ as a constrain to fix the initial conditions and the viscosity parameter of viscous evolutionary models, we use approximate Bayesian modeling
to infer the dust properties that produce the observed decrease of the IR excess with age, in the range between $4.5\,\mu$m and $24\,\mu$m.
We calculate an extensive grid of irradiated disk models with a two-layered wall to emulate a curved dust  inner edge and obtain the vertical structure consistent with the surface density predicted by viscous evolution.
We find that the median dust depletion in the disk upper layers is $\epsilon \sim 3 \times 10^{-3}$ at 1.5 Myr, consistent with previous studies, and it decreases to
$\epsilon \sim 3 \times 10^{-4}$ by 7.5 Myr.
We include photoevaporation in a simple model of the disk evolution and find that a photoevaporative wind mass-loss rate of
$\sim 1 -3  \times 10 ^{-9} \, M_{\odot}yr^{-1}$ agrees with
the decrease of the disk fraction with age reasonably well.
The models show the inward evolution of the \H2O and CO snowlines.

\end{abstract}

\keywords{accretion, accretion disks ---
circumstellar matter --- stars: pre-main sequence --- stars: variables: T Tauri, Herbig Ae/Be}

\section{Introduction}
\label{sec_intro}

Protoplanetary disks are natural byproducts of the star formation process, and the sites where new planetary systems will emerge
\citep{Williams2011, Testi2014}.
Stars surrounded by protoplanetary disks exhibit 
excess flux over the photosphere from the infrared to millimeter wavelengths 
due to emission by dust particles; thus the spectral energy distribution (SED) of the  system is directly related to the structure and the physical conditions of the dusty component of the disk
\citep{Dalessio_1998, Dalessio_2001,Dalessio_2006}.
In addition, disks are accreting mass onto the star and
releasing accretion energy; this accretion
energy appears as a continuum excess that veils the photospheric lines and dominates in the UV continuum and in emission lines 
\citep{Calvet_1998,Hartmann_2016}.

Protoplanetary disks show
direct evidence of evolution. 
Studies of disks
in populations ranging from $\sim$ 1 to 10 Myr indicate that
the number of disk-bearing stars decreases with age, from a 
frequency of $\sim 70$ \% at 1 Myr to $<$ 10\% by 10 Myr
\citep{Hernandez_2007a, Hernandez_2008, Briceno_2019}. The frequency of
stars accreting mass from their disk also decreases with age,
in proportions similar,
but not equal, 
to
the frequency of inner disks 
estimated from dust emission
\citep{Fedele_2010,Briceno_2019}.

It has also been found that the mass accretion rate decreases with age
\citep[][and references therein]{Hartmann_1998,Hartmann_2016}, which is
consistent with the viscous evolution of  
self-similar disks 
\citep{LP_1974, Hartmann_1998}.

In addition,
it has also been found that
the median strength of the
near-infrared excess in
disks of a given population
decreases 
as the age of the population 
increases \citep{Hernandez_2007b,Luhman_2010}. 
The IR excess arises from emission from dust in the disk,
which initially constitutes about 1\% of the total mass.
The observed decrease of the strength of the excess
may be consistent with the decrease of total disk mass
with time resulting from viscous evolution. Alternatively,
in addition to gas evolution, the decay of emission may require
dust evolution, such as growth and settling towards the
midplane \citep{Weidenschilling_1997, Dullemond_2004, Birnstiel_2009, Birnstiel_2010, Birnstiel_2011}.
Despite recent progress, 
details of the processes leading to the  
decrease of dust emission with age are still 
uncertain.

Disk censuses using broad band photometric SEDs indicate that the 
decrease of IR emission 
leading to the evolution from optically thick full primordial disks to optically thin debris disks follows at least two evolutionary path-ways \citep{Hernandez_2007b, Merin_2010, Cieza_2007, Williams2011}. One 
is
represented by the {\it evolved disks}
in which the flux at all bands decreases with age
\citep{Hernandez_2008,Luhman_2010}, suggesting a gradual 
depletion
of dust at all disk radii \citep{Merin_2010}.
These disks are known as evolved (used here), ``anemic'', ``homologously depleted'',
and
``homogeneously draining'' \citep{Lada_2006,Currie_2009,Koepferl_2013}
and no large inner cavities are detectable in their SEDs. 

Another path-way of disk evolution  comes from observations of inner disk zones cleared of
small
dust, which appear as flux deficits in the near-IR
in the {\it transitional disks} (TDs). Observationally, classical TDs have been defined as protoplanetary disks with little or no near-IR excess ($\lambda \lesssim 10\mu m$) and significant excess comparable to the median of Taurus 
\citep{Mauco_2016,Furlan_2006} at longer wavelengths \citep{Strom_1989, Calvet_2002,Calvet_2005}.
Several explanations have been given for opening holes in the inner disk: dynamical clearing by planets, photoevaporation and viscous evolution, dust grain growth, differential dust drift, dead zones and condensations fronts
\citep{Zhu_2011,Zhu_2012,Espaillat_2014, Owen_2016,Alexander_2014,Chiang_2007,Dullemond_2005,Birnstiel_2012,Pinilla_2016,Zhang_2015}.
The {\it pre-transitional disks} \citep[PTDs][]{Espaillat_2007, Espaillat_2010}, those with an
optically thick inner ring inside their large
cavities, 
may correspond to a phase of evolution previous to TDs
in which the innermost disk has not yet completely dissipated. 
The most accepted explanation 
is that the large gaps 
are due to planets, 
which have already been found
in the cleared regions of
pre-transitional disks
\citep{Keppler_2019,Haffert_2019}. 

ALMA high angular resolution observations at submillimetric wavelengths have shown that
structure is ubiquitous in
protoplanetary disks, with multiple rings and spiral arms
detected in numerous disks
\citep{ALMA_2015,Andrews_2016, Andrews_2018}; TDs and PTDs are extremes cases in which the dust structure also appears in the SED.
Several interpretations have been suggested for
these structures. For instance,
large dust particles get trapped at the edges of
gaps opened by forming planets, resulting in bright rings
\citep{Zhang_2018}.
Alternatively,
the gaps could be due to changes in
the dust properties, as happens in
sublimation fronts
\citep{Zhang_2015}.
In any event,
observations of disk
structures are biased towards the bright and largest disks, those that can be resolved
\citep{Andrews_2018,Long_2018,Long_2019}.
On the other hand, 
except for the TDs, disks with detected structure share the same 
location in near and mid-IR color-color diagrams than unresolved disks
\citep[cf. IRAC colors in][]
{Hartmann_2005},
which suggests
a common
overall evolution, at least at the inner regions.

Several studies give different estimates for the relative numbers of (pre)transitional and evolved disks in populations of different ages. 
Part of these differences arises because the  
criteria to define a transitional or an evolved disk are far from homogeneous 
\citep{Williams2011}. 
In this contribution we use the
observational definition of TD as given above, which physically corresponds to disks  with inner holes of tens of au
\citep{Espaillat_2012, Espaillat_2014}. This kind of disks represent $<$10-20\%  of the total disk population in young stellar clusters \citep{Muzerolle_2010}. 
Our focus in this work will be the rest of the disks, which encompass the 
``weak'' and ``warm'' excess disks of
\citet{Muzerolle_2010},
the 
``primordial ultra-settled'' and the ``homogeneous draining'' of 
\citet{Koepferl_2013} and \citet{Ercolano_2015},
the ``anemic'' of \citet{Currie_2009},
and the 
``evolved'' of
\citet{Hernandez_2008} and
\citet{Luhman_2010}.

We carry out a new disks census including well known nearby  young  stellar  groups  studied  using  IRAC   photometry \citep[3.6, 4.5, 5.8 and 8 $\mu m$;][]{Fazio_2004} and the 24 $\mu m$ MIPS photometry \citep{Rieke_2004}. In general, we also use optical PANSTARRS data \citep{Chambers_2016} to trace the stellar photosphere of the disk-bearing stars. These groups have ages between 1 and 11 Myr, 
the critical age range where significant
disk evolution takes place
\citep{Hernandez_2007a}.
We study the evolution of dust
using a new indicator of
the flux excess over the photosphere
in the near and mid-IR, between 4.5 $\mu m$ and 24 $\mu m$. We also use the D'Alessio irradiated accretion disk (DIAD) models to infer the state of the dust at a given age using approximate Bayessian modeling, and in particular, we focus on
indicators of dust settling at 
two size scales, at $\sim$ 0.1 au and at $\sim$ 10 au, to explore the degree of inner disk clearing. In addition, in contrast to previous works, we include the decrease of the observed mass accretion rate with age as an additional constraint in the evolution.

The structure of this paper is
the following.
In Section \ref{sec_observed_disk_ev}, we present 
the general properties of the stellar populations studied in this work,
calculate disks excesses and mass
accretion rates, and discuss
how they change with age.
In Section \ref{sec_Interpretation of observed evolution} we 
describe
the disk models 
that we use
to interpret the observations,
and the results
inferred from using these
models to interpret the
disk evolution indicators. 
In Section \ref{sec_discussion} we discuss our results and put them in context of what we know from previous studies. Finally, in Section \ref{sec_summary_conclusions} we summarize and present our conclusions.

\section{Observed disk evolution}
\label{sec_observed_disk_ev}

\subsection{Sample of stellar populations}
\label{sec_sample}
 
To make a comprehensive statistical study of the evolution of disks around T Tauri stars (TTSs), we selected nearby ($\lesssim 500$ pc), young ($\sim 1$ to 11 Myr old) stellar groups with Spitzer photometry 
and with available spectroscopic information for most of the
TTSs reported in each stellar group. The populations included in
this work are (see Table \ref{tab_groups}): the Orion Nebula Cluster (ONC), Taurus, the IC348 cluster, the $\sigma$ Orionis cluster ($\sigma$ Ori), the $\lambda$ Orionis cluster ($\lambda$ Ori),  the Orion OB1b subassociation (Ori OB1b), the Upper Scorpius subassociation (UpSco), the $\gamma$ Velorum cluster ( $\gamma$ Vel), and the Orion OB1a subassociation (Ori OB1a), which includes the stellar aggregates 25 Ori, HD 35762, and HR 1833.

For each group, we compiled stars with spectral types ranging from K0-M6. To  study the evolution of the dust component of the disk (\S\ref{subsec_DCE} \& \ref{subsubsec_results_DCE_dist}), 
we only selected 
stars with 
additonal IRAC photometry in all four bands. In \S \ref{subsec_DCE} we describe the methodology followed to separate the disk-bearing stars from the rest, in each stellar group. To study the gas component of the disk (\S\ref{subsec_gas_evolution}), we use a different sample from the one used to study the dust component, since in this case we selected confirmed
TTSs  
with 
existing equivalent width of {H$\alpha$} (EW {H$\alpha$}), measured using the SpTClass tool \citep{Hernandez_2004,Hernandez_2017}, and with GAIA DR2 parallaxes with relative errors $\lesssim 20\%$, regardless of whether or not they have IRAC photometry.  
This means that some stars in the sample used to calculate mass accretion rates additionally have IRAC photometry and thus they also belong to the sample used to study the dust component. However there are stars which do not have IRAC photometry, while having the necessary information to determine the corresponding mass accretion rate. As a consequence, there are stars in this sample of accretors, which do not belong to the sample used to study the dust component using color excesses and accordingly, there are stars with color excesses for which $\dot{M}$ could not be determined. 

In general, we use the color $[g-i]$ from PANSTARRS \citep{Chambers_2016} to estimate homogeneously visual extinctions for the samples of TTSs (see \S \ref{sec_av}). Thus we also require that the TTSs have $g$ and $i$ magnitudes reported by PANSTARRS \footnote{Except for the $\gamma$ Velorum cluster, which was not observed by the PANSTARRS survey}. Table \ref{tab_groups} shows the stellar populations included in this work with their general properties taken from the literature, except for the last four columns. Based on both the stellar ages and disk frequencies previously reported, we grouped the stellar population into four age bins (see last column of Table \ref{tab_groups}).\\

\begin{deluxetable*}{lcccccccc}
	\tablecaption{Stellar Groups
	\label{tab_groups}}
\tabletypesize{\scriptsize}
\tablehead{
	\colhead{Group} &
	\colhead{Mean distance }  & 
	\colhead{Age} &  
	\colhead{Disk fraction} & 
	\colhead{References} &
	\colhead{Initial} & 
	\colhead{Disk-bearing} &
	\colhead{Accretors} &
	\colhead{Age}\\
	\colhead{} & 
	\colhead{pc }  & 
	\colhead{Myr} &  
	\colhead{\%} & 
	\colhead{ }  & 
	\colhead{sample} &
	\colhead{stars} & 
	\colhead{sample} &
	\colhead{bin} 
}
\startdata
ONC         & 400     & 1-3 & 73$\pm$5.9 & 4,7,4 & 508 & 305 & 228 & 1\\ 
Taurus      & 130-200 & 1-2 & 63.7$\pm$5.1 & 8,9,* & 137 & 87 & 16  & 1\\ 
IC348       & 320     & 2-3 & 47$\pm$12 & 10,1,1 & 163 & 59 & \nodata & 2\\ 
$\sigma$ Ori & 400    & 3   & 36$\pm$4  & 11,2,2 & 185 & 73 & 40  & 2\\ 
$\lambda$ Ori & 400   & 4-6 & 18.5$\pm4$ & 12,3,3 & 142 & 24 & \nodata & 3\\ 
Ori OB1b      & 400  & 5 & 13-17 & 4,6,4 & 278 & 45 & 59  & 3\\ 
Upper Sco.    & 146  & 5-11 & 25$\pm$0.02 & 13,14,9 & 59  & 19 & \nodata & 3\\ 
$\gamma$ Vel & 345   & 7.5  & 5-7  & 15,16,5 & 125 & 14 & \nodata  & 4\\ 
Ori OB1a   & 350-360 & 9-14 & 6-10 & 4,4,6 & 189 & 13 & 77 & 4\\ 
\enddata
 \tablecomments{References: 
(1) \citet{Lada_2006}, (2) \citet{Hernandez_2007a}, (3) \citet{Hernandez_2010}, (4) \citet{Briceno_2019}, (5) \citet{Hernandez_2008}, (6)  \citet{Hernandez_2007b}, (7) \citet{Hillenbrand_2013}, (8) \citet{Galli_2019}, (9) \citet{Luhman_2012}, (10) \citet{Ortiz_2018}, (11) \citet{Perez_2018}, (12) \citet{Kounkel_2018}, (13) \citet{Galli_2018}, (14) \citet{David_2019}, (15) \citet{Franciosini_2018}, (16) \citet{Jeffries_2017}. *Using the data from \citet{Luhman_2010} we estimate a disk fraction of $63.7\pm5.1$ in Taurus.}
\tablecomments{Column 6 refers to the number of K0-M6 stars initially compiled (diskless and disk-bearing stars) with IRAC and PANSTARRS photometry in each stellar group. Column 7 refers to the number of stars with near-IR excess above the photosphere, separated according to \S \ref{subsec_DCE}, which correspond to the sample used to study the dust component of the disks. Column 8 refers to the CTTSs/CWTTSs sample, described in \S \ref{subsubsec_observed_mdots} used to study the evolution of $\dot{M}$.}
\end{deluxetable*}

\subsubsection{ONC}
\label{sec_onc}
The ONC is a very young stellar cluster \citep[$\sim 1-3$ Myr;][]{Hillenbrand_1997,Hillenbrand_2013,DaRio_2010, Megeath_2016} with a mean distance of $\sim$ 400 pc \citep[][using GAIA DR2 data]{Briceno_2019}. Also \citet{Briceno_2019} estimate a disk frequency of $\sim 73\pm 5.9\%$. The initial sample of TTSs includes 782 stars with spectral types, equivalent widths of \ion{Li}{1} $\lambda$6708\AA (EW LiI), and EW {H$\alpha$} obtained from the analysis of low-resolution spectra observed with the fiber-fed multi-object Hectospec instrument mounted on the 6.5 m Telescope of the MMT Observatory (Hern\'andez et al. 2020a, in preparation). We also include 127 TTSs with spectral types, EW LiI, and EW {H$\alpha$} reported by \citet{Briceno_2019}. Spectral types, EW LiI, and EW {H$\alpha$} in Hern\'andez et al. 2020a (in preparation) and \citet{Briceno_2019} were obtained using the SpTClass code \citep{Hernandez_2017}, an IRAF/IDL code based on the methods described in \citet{Hernandez_2004}. We complete the initial sample adding 86 TTSs and 48 TTSs with spectral types reported by \citet{Hillenbrand_1997} and \citet{Hillenbrand_2013}, respectively. Out of 1036 TTSs in the initial sample, 500 TTSs have IRAC/MIPS photometry reported by \citet{Megeath_2012}. We also include 170 additional TTSs with IRAC/MIPS photometry provided by Tom Megeath (private communication) which were not included in \citet{Megeath_2012}. Finally, out of 670 TTSs with IRAC/MIPS photometry, we select 508 TTSs with $g$ and $i$ magnitudes from PANSTARRS .\\

\subsubsection{Taurus}
\label{sec_taurus}
Taurus is one of the best studied and closest star-forming molecular cloud complexes, 
 with an estimated age of $\sim 1-2$ Myr \citep{Luhman_2010, Furlan_2011}. Using data from \citet{Luhman_2010} we estimate a disk frequency of this complex of $\sim 63.7\pm5.1\%$ for K0-M6 stars. Based on distances provided by GAIA DR2 and Very
Long Baseline Interferometry (VLBI) observations, \citet{Galli_2019} show that the Taurus star forming complex includes several molecular clouds 
located at different distances (from $\sim$130 pc to $\sim$200 pc). Based mostly on the work of \citet{Hartmann_2005} and \citet{Luhman_2010}, \citet{Esplin_2014} compiled 414 members of Taurus with IRAC/MIPS photometry.  From this sample, we selected 260 TTSs with measurements in all the photometric bands of IRAC and with spectral types ranging from K0 to M6. In some cases,  stars identified as
binary stars have spectral types for the two components (e.g. M1+M7); in this case we select the spectral type of the brighter
component (e.g. M1). Out of 260 TTSs, 137 TTSs have $g$ and $i$ photometry from PANSTARRS. A sample of 100 TTSs have optical low resolution spectra available from the FAST Public Archive \footnote{http://tdc-www.harvard.edu/cgi-bin/arc/fsearch}. We have measured EW {H$\alpha$} in this work for this sample using the SpTClass tool \citep{Hernandez_2017}. \\

\subsubsection{IC348}
\label{sec_ic348}
IC348 is a nearby and compact young cluster ($\sim$ 2-3 Myr) located in the Perseus OB2 star forming region \citep{Lada_2006,Herbst_2008}. Combining trigonometric parallaxes from 
Very Long Baseline Array (VLBA) observations and GAIA DR2 parallaxes, \citet{Ortiz_2018} estimate a distance of $\sim$320 pc for this cluster. \citet{Lada_2006} estimated a disk frequency of $\sim47\pm 12\%$ and provide spectral types and IRAC/MIPS photometry for 307 stars in the IC348 cluster. From this set,  
we selected 220 stars with spectral types ranging from K0 to M6 and with photometric measurements in all IRAC bands. Out of 220 TTSs, 163 TTSs have PANSTARRS $g$ and $i$ magnitudes. \\

\subsubsection{$\sigma$ Ori}
\label{sec_sori}
$\sigma$ Ori is a young ($\sim$ 3 Myr) and relatively populous stellar cluster located in the Orion OB1b sub association and with a disk frequency of $\sim 36\pm 4\%$ \citep{Hernandez_2007a}. Using GAIA DR2 parallaxes, \citet{Perez_2018} estimate a mean distance of $\sim$400 pc. We have selected 221 TTSs with spectral types ranging from K0 to M6, EW LiI, and EW {H$\alpha$} obtained by \citet{Hernandez_2014} using the SpTClass tool and with photometric measurements in all IRAC bands \citep{Hernandez_2007a}.  From this sample, 185 TTSs  have PANSTARRS $g$ and $i$ magnitudes.\\

\subsubsection{$\lambda$ Ori}
\label{sec_lori}

$\lambda$ Ori is a young ($\sim$4-6 Myr) and relatively populous stellar group with an overall disk frequency for M type stars of $\sim 18.5\pm4\%$ \citep{Hernandez_2010}. Combining spectroscopic and astrometric data from APOGEE-2 and GAIA DR2, \citet{Kounkel_2018} reported an average distance of $\sim$400 pc for this stellar cluster. Spectral types were compiled from \citet{Bayo_2011} and \citet{Sacco_2008}. We also add stars with effective temperatures reported by \citet{Bayo_2008} and \citet{Bayo_2011}. Those effective temperatures were converted to spectral types using Table 6 from \citet{PM_2013}. From this compilation we have selected 181 TTSs with spectral types ranging from K0 to M6 and with photometric measurements in all IRAC bands \citep{Hernandez_2010}. From this sample, we select 142 TTSs with PANSTARRS $g$ and $i$ magnitudes.\\

\subsubsection{Ori OB1b}
\label{sec_oriOB1b}
\citet{Briceno_2019} report EW LiI and EW {H$\alpha$} for 551 TTSs with spectral types ranging from K0 to M6 located in this subassociation. Based on GAIA DR2 parallaxes and the PMS models of \citet{Siess_2000}, \citet{Briceno_2019} adopted a distance of 400 pc and a stellar age of $\sim$5 Myr. 
The disk frequency in Ori OB1b is $\sim 13-17 \%$ \citep{Hernandez_2007b,Briceno_2019}.
Out of 551 TTSs, 104 and 213 TTSs have photometric measurements in all IRAC bands provided by \citet{Hernandez_2007b} and Hern\'andez et al. 2020b (in preparation), respectively. From this sample, we select 278 TTSs with PANSTARRS $g$ and $i$ magnitudes. \\

\subsubsection{Upper Sco}
\label{sec_us}
Upper Sco  is the youngest 
stellar population of the Scorpius-Centaurus OB association, which is the nearest region of recent massive star formation \citep{Preibisch_2008}.
Based on the GAIA first data release, \citet{Galli_2018} estimate a mean distance of $\sim$146 pc. Stellar ages estimated for Upper Scorpius range from $\sim$5 \citep{Preibisch_2002} to 11 Myr \citep{Pecaut_2012} and recently, using eclipsing binaries, \citet{David_2019}  estimate and age of 5-7 Myr. Out of 306 stars in \citet{Luhman_2012} with  spectral types ranging from K0 to M6, 74 stars have protoplanetary disks. The disk frequency reported by \citet{Luhman_2012} is $\sim25\pm0.02\%$ for Upper Scorpius. Out of 306 K and M stars,  90 stars have photometric measurements in all IRAC bands. Finally, out of 90 stars, we select 59 stars with PANSTARRS $g$ and $i$ magnitudes.\\

\subsubsection{$\gamma$ Vel}
\label{sec_gvel}
$\gamma$ Vel is a young stellar cluster with a central binary system consisting of an O7.5 star and a Wolf-Rayet star \citep[WC8;][]{Hernandez_2008}. Based on GAIA DR2 observations, \citet{Franciosini_2018} estimate a distance of $\sim$345 pc. The stellar age of the $\gamma$ Velorum cluster can range from 5 Myr to 20 Myr, with an age of 7.5$\pm$ 1 Myr inferred from the color-magnitude diagram
\citep{Jeffries_2017}. The disk frequency of the cluster is $\sim 5-7\%$ \citep{Hernandez_2008}. Out of 557 stars with optical photometry (V, $\text{I}_{\text{c}}$) and IRAC photometry in all bands \citep{Hernandez_2008}, we compile 125 stars with effective temperatures
lower than 5000 K \citep[e.g. spectral type K0 or later;][]{PM_2013} reported by \citet{Spina_2014}, \citet{Frasca_2015} or \citet{Smiljanic_2016}. From the sample of 125 stars, 11 do not have reported spectral types or effective temperature, thus we estimate their stellar mass using the $[\text{V}-\text{I}_{\text{c}}]$ vs stellar mass relation from \citet{Prisinzano_2016}, then we estimate their effective temperature from the relation $T_{\text{eff}}=1005.78 \, \, M_{\ast}+3042.1$ \citep{Frasca_2015}. Since $\gamma$ Vel is out of the area coverage of the PANSTARRS survey, we estimate visual extinction using the $[\text{V}-\text{I}_{\text{c}}]$ color instead the $[g-i]$ color (see \S \ref{sec_av}).\\

\subsubsection{Ori OB1a}
\label{sec_oriOB1a}
\citet{Briceno_2019} report EW LiI and EW {H$\alpha$} for 1211 TTSs with spectral types ranging from K0 to M6 located in this sub-association. This sample includes 807 TTSs located in the dispersed population of Ori OB1a,  and 404 TTSs members of the stellar groups 25 Ori, HR 1833, and HD 35762. 
Since some TTSs of the dispersed population of Ori OB1a are located close to other younger stellar groups, we rejected 100 TTSs located near the $\lambda$ Ori star forming region (DEC$>$4$\degree$) or near the Ori OB1b subassociation (DEC$<$-2$\degree$).  Based on GAIA DR2 parallaxes and the PMS models of \citet{Siess_2000}, \citet{Briceno_2019} adopted a distance of 350-360 pc with a stellar age range of  $\sim$9-14 Myr. 
The disk frequency for the Orion OB1a subassociation is $\sim6-10\%$ \citep{Hernandez_2007b,Briceno_2019}.
Out of 1211 TTSs, 121 and 93 TTSs have photometric measurements in all IRAC bands, 
reported in \citet{Hernandez_2007b} and Hern\'andez et al. 2020b (in preparation), respectively. From this sample, we select 189 TTSs with PANSTARRS $g$ and $i$ magnitudes. \\

\subsection{Visual extinctions, stellar luminosities and stellar masses}
\label{sec_av}

In general, we calculate
the extinction \Av for each star in our samples, using the $[g-i]$ color from PANSTARRS. 
One of the advantages of using PANSTARRS photometry is that this survey was performed several times per filter, over a four-year time span, which addresses the effects of variability when using the reported mean magnitude.
The intrinsic photospheric color $[g-i]_{\text{o}}$ was added to Table 6 from \citet{PM_2013} using the following relation between the
color $[\text{V}-\text{I}_{\text{c}}]$ and the PANSTARRS color $[g-i]$:

\begin{equation}
[g-i]= 0.475(\pm 0.019) + 0.751(\pm 0.007) \, [\text{V}-\text{I}_{\text{c}}].
\label{eq_g-i}
\end{equation}

This relation was obtained using a sample of confirmed TTSs with both $[\text{V}-\text{I}_{\text{c}}]$ color reported by \citet{Briceno_2019} and $[g-i]$ color from PANSTARRS.\\

We use Table 6 from \citet{PM_2013} to interpolate the spectral types, and obtain the effective temperature $T_{\text{eff}}$, the bolometric correction for the $J$ band ($\text{BC}_{J}$), and the intrinsic color $[g-i]_{\text{o}}$. The extinction \Av is estimated
from the $[g-i]$ color
using the standard interstellar reddening law ($R_\text{V}$=3.1) and relations from \citet{Cardelli_1989}, with a central wavelength of 0.481 {\micron} and 0.752 {\micron} for the $g$ and $i$ band, respectively \citep{Tonry_2012}. \\

Since PANSTARRS does not cover the region of $\gamma$ Vel, the extinction \Av for this stellar cluster is estimated from the color
$[\text{V}-\text{I}_{\text{c}}]$ and the standar
interstellar reddening law, and the relation from \citet{Cardelli_1989}, with a central wavelength of 0.79 {\micron} for the $\text{I}_{\text{c}}$ filter. The intrinsic color $[\text{V}-\text{I}_{\text{c}}]_{\text{o}}$ was estimated interpolating the effective temperature in Table 6 from \citet{PM_2013}. 

We estimate stellar luminosities for TTSs with uncertainties in parallaxes below 20\%
\citep{GAIA_2018}. Thus, distances can be calculated as the inverse of the parallaxes 
\citep{Bailer_2015}. 
We derive the absolute
bolometric magnitude from the
absolute $J$ magnitude and the
bolometric correction 
$\text{BC}_{J}$ from 
\citet{PM_2013}. The 
2MASS $J$ magnitudes \citep{Cutri_2003}
were corrected for 
extinction using the
derived \Av and the
standard reddening law.
Finally,  the
luminosities are obtained using 
$M_{bol} = 4.74 $ for the Sun.

The stellar mass ($M_{\ast}$) is obtained by performing a double interpolation in the H-R diagram using the Pre-Main Sequence (PMS) models of \citet{Siess_2000}. Some stars have luminosity or effective temperature out of the range of these stellar evolutionary models, hence their mass cannot be determined.  

\subsection{Disk color excess}
\label{subsec_DCE}

We 
introduce
the disk color excess (hereafter DCE) as  
a new
indicator of disk infrared excesses  to gauge the global dust evolution. This indicator takes into account not only the extinction correction of infrared colors used as disk tracers, but also 
the photospheric contribution for a given spectral type. We use the $J$ band 
as  
anchor of the colors to measure the infrared excesses over the photosphere because this band is representative of the stellar photosphere of TTSs,
since it
 is less contaminated by infrared excesses produced by the dust component in the disk 
 or
 by excess emission from accretion shocks than other optical-red bands. 
For each star in each population described in \S\ref{sec_sample}, we
define the DCE for 
each Spitzer color 
[$J-\lambda_{IR}$]
as:

\begin{equation}
\text{DCE}_{J-\lambda_{IR}}=[J-\lambda_{IR}]_{\text{obs}}-[J-\lambda_{IR}]_{\text{o}}-\left(\frac{A_{\text{J}}}{A_{\text{V}}}-\frac{A_{\lambda_{IR}}}{A_{\text{V}}}\right)\times A_{\text{V}},
\label{eq_EDE}
\end{equation}

\noindent
where $\lambda_{IR}$ represents the Spitzer magnitudes [3.6], [4.5], [5.8], [8.0] and [24], and $[J-\lambda_{IR}]_{\text{o}}$ represents the intrinsic Spitzer photospheric colors. 
The selective extinctions, $A_{3.6}/A_{\text{V}}$=0.048, $A_{4.5}/A_{\text{V}}$=0.035, $A_{5.8}/A_{\text{V}}$=0.024, $A_{8.0}/A_{\text{V}}$=0.031, and $A_{24}/A_{\text{V}}$=0.015, are estimated interpolating the central wavelengths of each filter using Table 21.6 from Allen's Astrophysical Quantities \citep{Cox_2000} assuming a standard interstellar law.
We estimate
 the photospheric Spitzer colors, $[J-3.6]_{\text{o}}$, $[J-4.5]_{\text{o}}$, $[J-5.8]_{\text{o}}$, $[J-8.0]_{\text{o}}$ and $[J-24]_{\text{o}}$, 
 by interpolating the spectral types in Table 13 from \citet{Luhman_2010}. For K-type stars earlier than K4, we assume the photospheric Spitzer colors of a K4 star. This approximation is in agreement within 1\% with empirical photospheric colors in the spectral type range K0-K4 (K. Luhman internal communication). At each age bin, the observational sample contains at most 3 K0-K4 stars (cf. Figure \ref{fig_all_bins_spectraltypes}), thus the estimated Spitzer colors are not affected, since the total sample used in each age bin is much larger than the number of K0-K4 stars.

\begin{figure}[h!]
\centering
\includegraphics[width=0.8\linewidth]{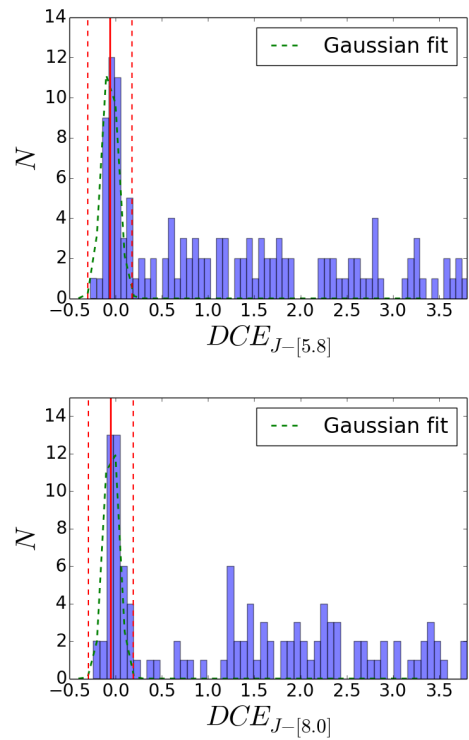}
\caption{\small $DCE_{J-[5.8]}$ distribution (upper panel) and $DCE_{J-[8.0]}$ distribution (lower panel) for stars in the Taurus sample. The green line is the Gaussian fit to the distribution of DCEs and the red lines are the mean (solid line) and the $3\sigma$ limits (dashed lines).
}
\label{DCE_Taurus}
\end{figure}

For each stellar group, we separate the disk-bearing stars from the diskless stars using the distributions of $DCE_{J-[5.8]}$ and $DCE_{J-[8.0]}$. As an illustration, Figure \ref{DCE_Taurus} shows these distributions for Taurus. We estimate the photospheric region by fitting a Gaussian function (green dashed line) to the DCE distributions around the photospheric values (e.g. DCE$\sim 0$).  We apply the $3\sigma$-clipping method, which consists of calculating the center value ($C_G$) and the standard deviation ($\sigma_G$) of the Gaussian function and redo the Gaussian fitting for stars within the range $C_G-3\times\sigma_G<DCE<C_G+3\times\sigma_G$. This process is repeated until the values for the center and $\sigma$ of the Gaussian function converge. In our populations, these values converge in less than 4 iterations. The final value of $C_g$ + $3\times\sigma_G$ is used to set the limit that separates disk-bearing stars from those with photospheric values. To avoid stars with  evolutionary stages earlier than Class II objects (e.g protostars), the upper limits for disk-bearing stars are set by estimating the DCEs associated to a flat SED ($DCE_{J-[5.8]}$=4.28; $DCE_{J-[8.0]}$=5.30). 
Thus, the disk-bearing stars sample is defined as those stars between the photospheric limit and the flat SED limit. The number 
of
TTSs selected as disk-bearing stars in each stellar population is indicated in Table \ref{tab_groups} (column 7). \\

\begin{figure*}[h!]
\centering
\includegraphics[width=0.9\linewidth]{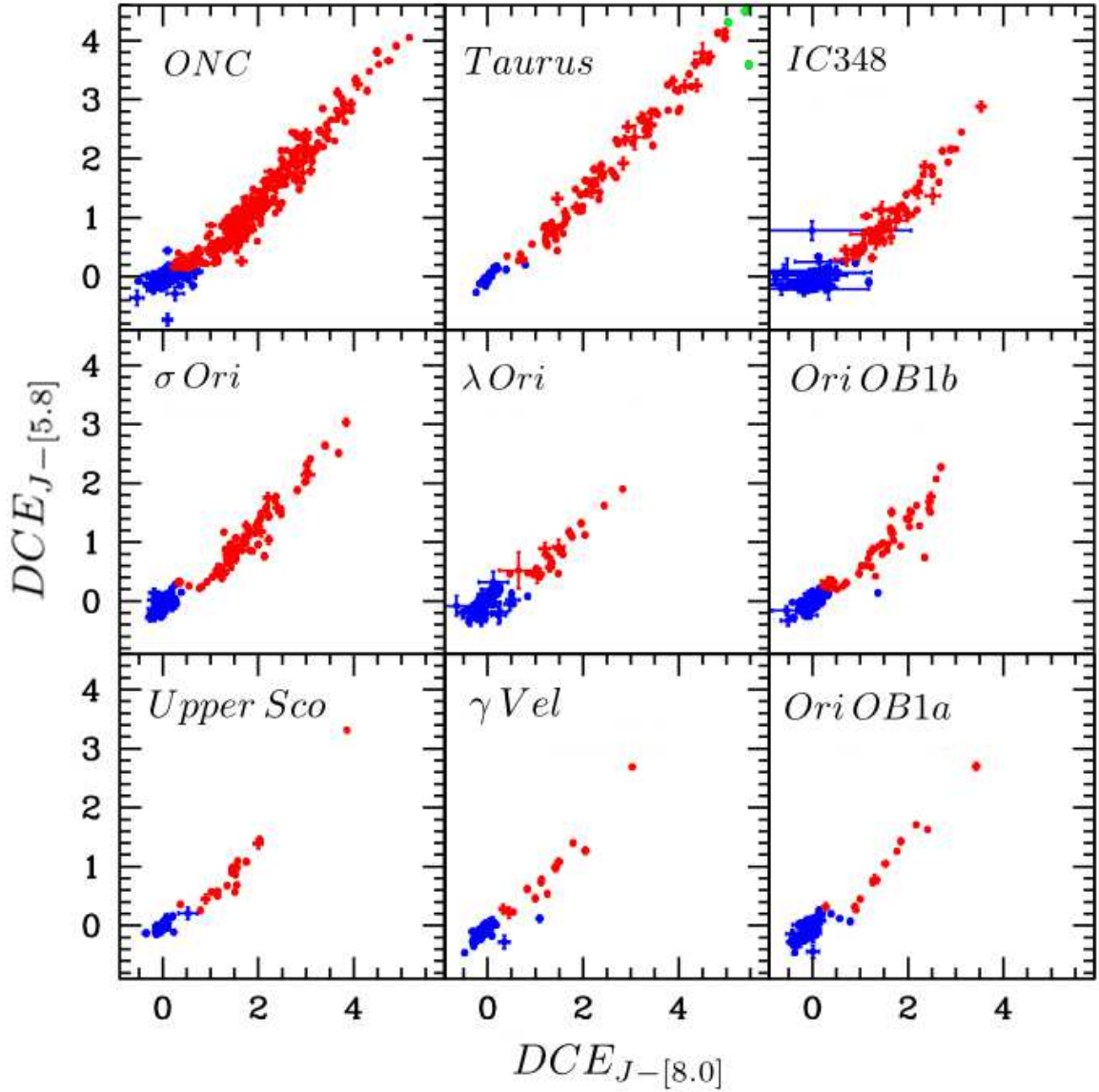}
\caption{\small $DCE_{J-[5.8]}$ vs $DCE_{J-[8.0]}$ diagrams for all stellar groups. Red dots correspond to the stars with excess at $5.8 \, \mu m$ and $8 \, \mu m$ and blue dots are stars without excess. The green points in the Taurus panel are objects in early stages than Class II objects.}
\label{fig_DCEs_1}
\end{figure*}

 Figure \ref{fig_DCEs_1} shows 
 the $DCE_{J-[5.8]}$ 
 color versus 
 the $DCE_{J-[8.0]}$
 color 
 for the stellar groups described in section \ref{sec_sample}, highlighting  the disk-bearing sample (red) selected to study the global evolution of the dusty component of the disk. The green points in the upper right corner in the Taurus panel correspond to stars in  evolutionary stages earlier than Class II objects. Following equation (\ref{eq_EDE}) we calculate, besides the DCEs$_{J-[5.8]}$ and DCE$_{J-[8.0]}$,  the rest of the DCEs ($J$-[3.6], $J$-[4.5] 
 and $J$-[24]) for the disk-bearing sample. Since the MIPS photometry is less deep than the IRAC observations and 
 the MIPS coverage can be slightly different  
 from the
 IRAC coverage, some disk-bearing stars do not have an
 estimation of $DCE_{J-[24]}$.  This bias can affect mainly 
 the
 disk-bearing stars with the smallest values of $DCE_{J-[5.8]}$ and $DCE_{J-[8.0]}$. As an illustration, 
Figure \ref{fig_all_bins_spectraltypes} shows the distribution of spectral types for all the disks-bearing stars (purple) and for the stars with MIPS 24 detections (pink) in all age bins (see \S \ref{subsec_separation}). The main difference in these distributions can be observed for stars with the latest spectral types.\\

\subsection{Separation into age bins}
\label{subsec_separation}

Since our goal is to make a
study with good statistical significance, we grouped the disk-bearing sample selected in each stellar population into 4 groups or age bins. 
We take into account two criteria to group the disk-bearing sample into the four age bins: their nominal age, obtained by fitting one isochrone to the location of all the stars of a given stellar group in the H-R diagram, and 
their disk fraction (see Table \ref{tab_groups}).
Following these criteria, we separate the
populations into  the following age bins:

\begin{itemize}
\item Bin 1 (1-2 Myr): ONC and Taurus
\item Bin 2 (2-3 Myr): IC348 and $\sigma$-Ori 
\item Bin 3 (3-5 Myr): $\lambda$ Ori, Ori OB1b, and Upper Sco
\item Bin 4 (5-11 Myr): $\gamma$ Vel and Ori OB1a 
\end{itemize}

\begin{figure}[h!]
\centering
\includegraphics[width=1\linewidth]{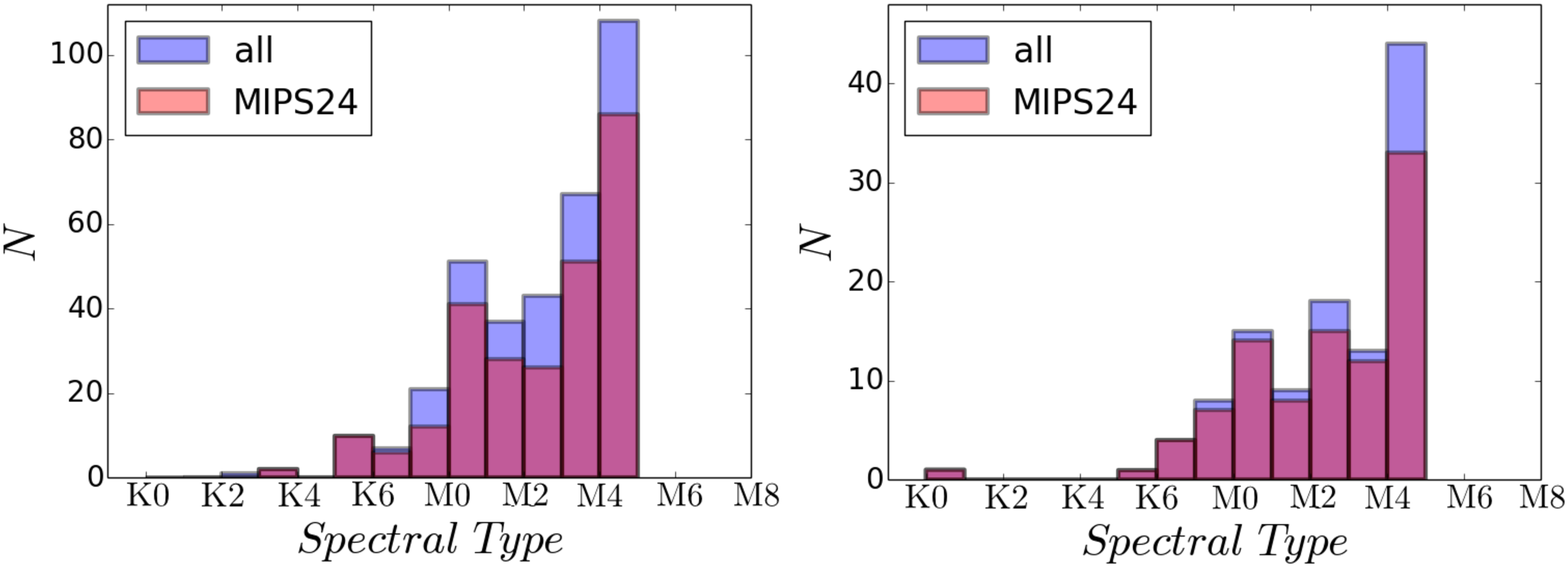}
\includegraphics[width=1\linewidth]{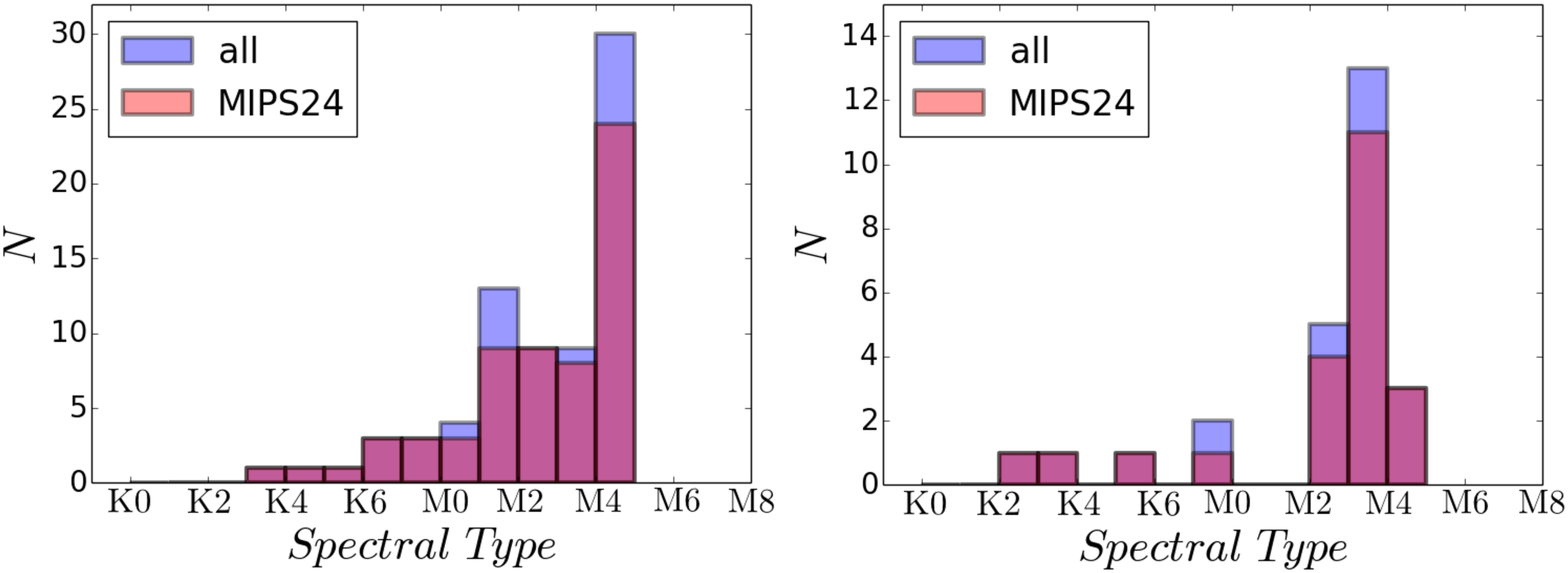}
\caption{\small  Distribution of spectral types at 1.5 Myr (bin 1, upper left), 2.5 Myr (bin 2, upper right), 4.5 Myr (bin 3, lower left) and  7.5 Myr (bin 4, lower right). Bin 1, 2, 3 and 4 comprehend 1-2, 2-3, 3-5 and 5-11 Myr, respectively. The purple distribution comes from all the stars in our sample, while the pink one corresponds to stars with MIPS 24 detections.}
\label{fig_all_bins_spectraltypes}
\end{figure}

\begin{figure}[h!]
\centering
\includegraphics[width=0.9\linewidth]{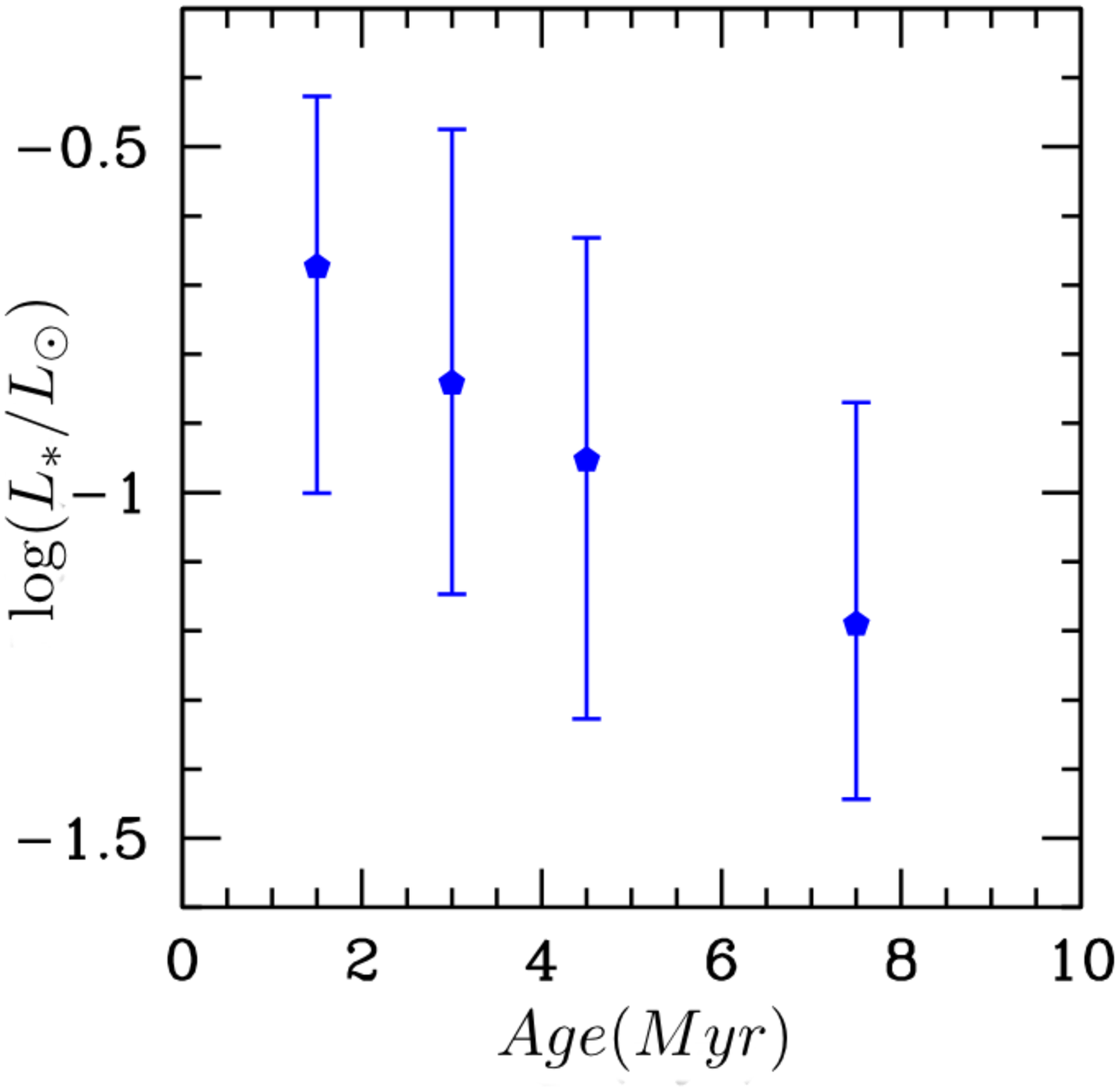}
\caption{\small  Median of the stellar luminosity (blue pentagons) for the stars in each age bin vs the age of the bin. The vertical bars represent the second and third quartiles.}
\label{fig_median_logLum_vs_age_susana}
\end{figure}

Table \ref{tab_groups} shows properties of all stellar populations in  \S \ref{sec_sample} and the age bin we assigned 
to them.
We assigned Upper Scorpius to the third age bin, in agreement with its  disk fraction and with the age (5-7 Myr) recently determined by \citet{David_2019}, although some authors suggest it is older 
\citep[$\sim$ 11 Myr,][]
{Pecaut_2012,Song_2012}.
To test the goodness of our 
assemblage of age bins, we plot the median of the stellar luminosity estimated in \S\ref{sec_av} vs the age of each bin (Figure \ref{fig_median_logLum_vs_age_susana}). It is apparent that the median of the stellar luminosity decreases with age, consistent with expectations from stellar evolution, since most stars in each bin are late K and M as shown in Figure \ref{fig_all_bins_spectraltypes}, and thus
are evolving along Hayashi tracks.

\begin{figure}[h!]
\centering
\includegraphics[width=1\linewidth]{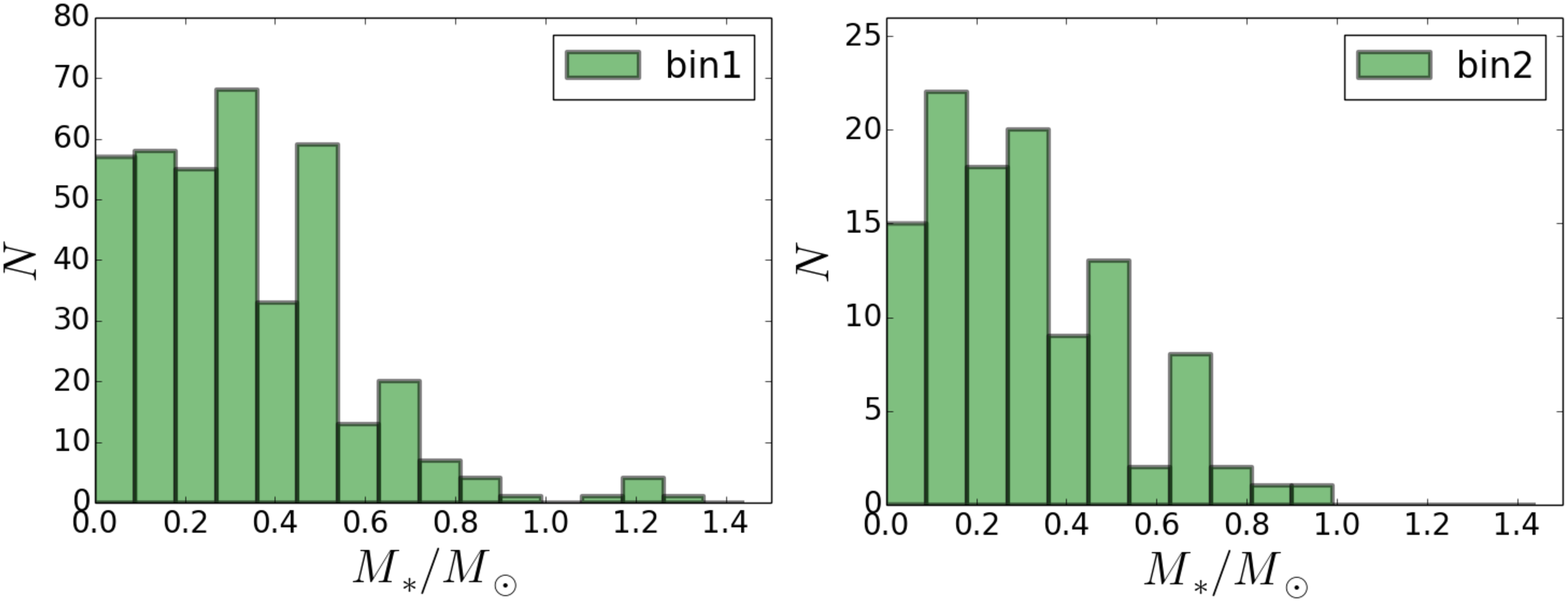}
\includegraphics[width=1\linewidth]{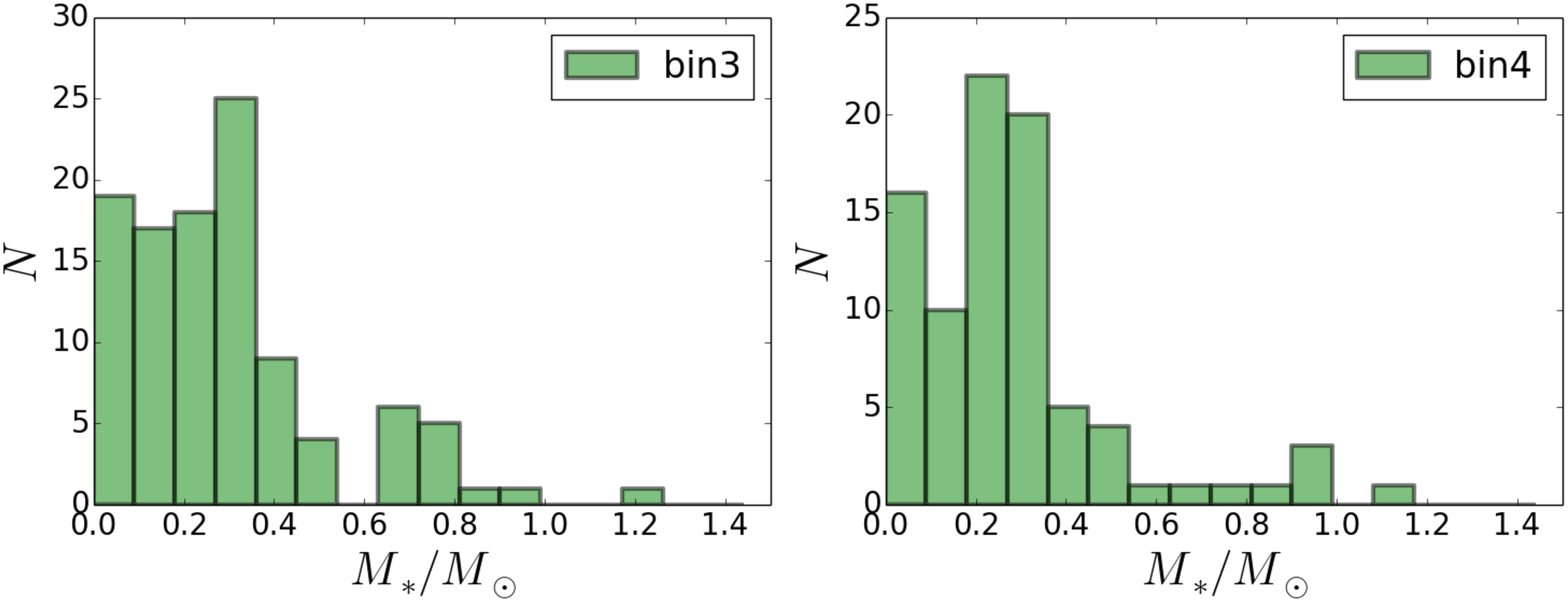}
\caption{\small Distribution of stellar masses in all bins.}
\label{fig_observed_masses_2}
\end{figure}

Figure \ref{fig_observed_masses_2} shows the distribution of stellar masses per age, including stars with determinations of mass accretion rates (\S \ref{subsubsec_observed_mdots}). The mean value of the observed stellar masses in all age bins is $\sim 0.3 \, M_{\odot}$ (see Table \ref{tab_stellar_masses}).

\begin{deluxetable}{lcccc}
    \tablecaption{Observed stellar masses
    \label{tab_stellar_masses}}
\tabletypesize{\scriptsize}
\tablehead{
    \colhead{bin} &
    \colhead{mean ($M_{\ast}/M_{\odot}$)} &
    \colhead{$\sigma$}
}
\startdata
1 & 0.33 & 0.24 \\
2 & 0.30 & 0.20 \\
3 & 0.29 & 0.22 \\
4 & 0.28 & 0.22 \\
\enddata
\end{deluxetable}

\subsection{The evolution of the disk color excess}
\label{subsubsec_results_DCE_dist}

\begin{figure*}[h!]
\centering
\includegraphics[width=1\linewidth]{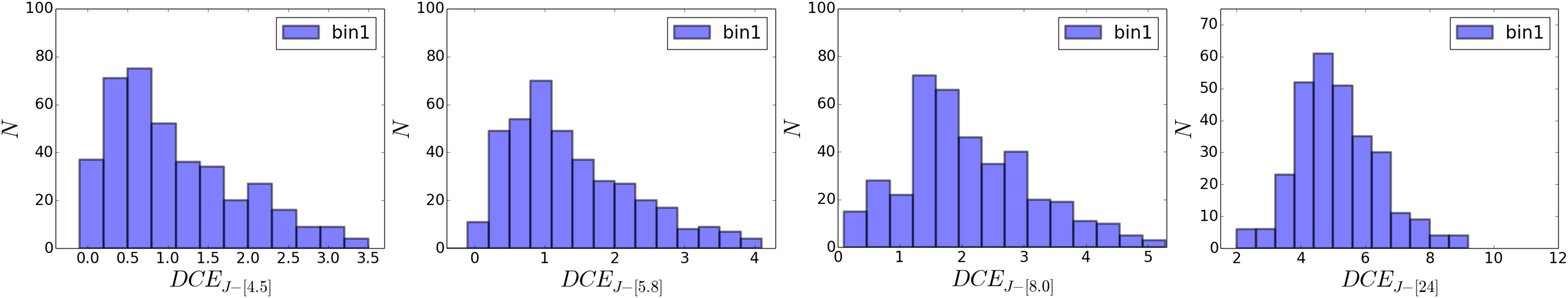}
\includegraphics[width=1\linewidth]{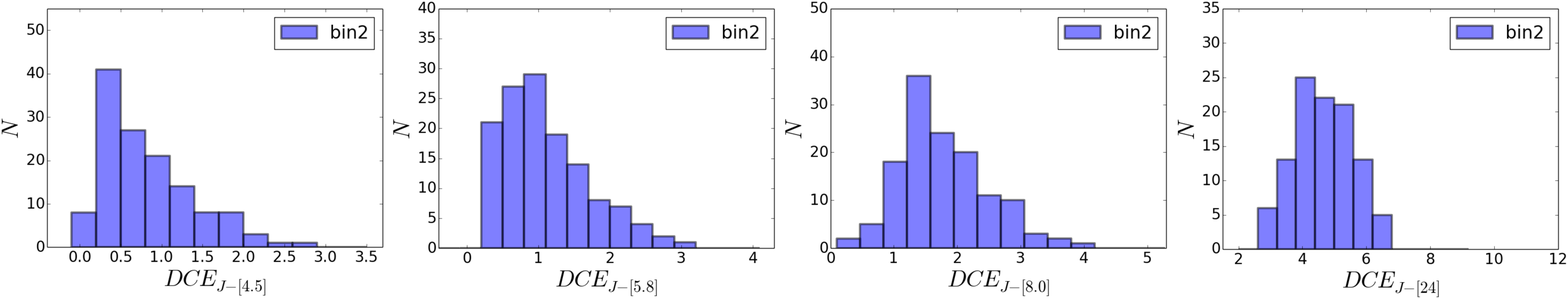}
\includegraphics[width=1\linewidth]{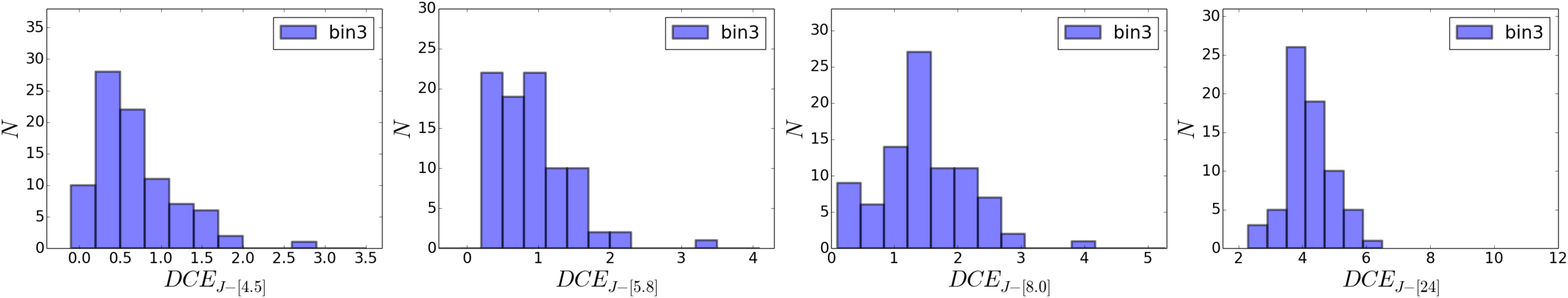}
\includegraphics[width=1\linewidth]{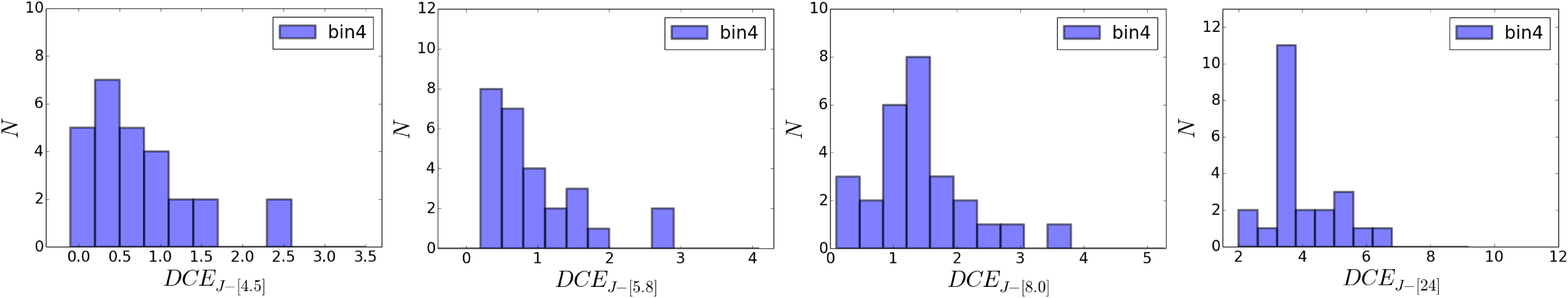}
\caption{\small DCE distributions for: J-[4.5](first column), J-[5.8](second column), J-[8.0](third column) and J-[24](fourth column). The age of the bins increases from top to bottom.} 
\label{fig_all_DCEs_all_bins}
\end{figure*}

\begin{figure}[h!]
\centering
\includegraphics[width=1\linewidth]{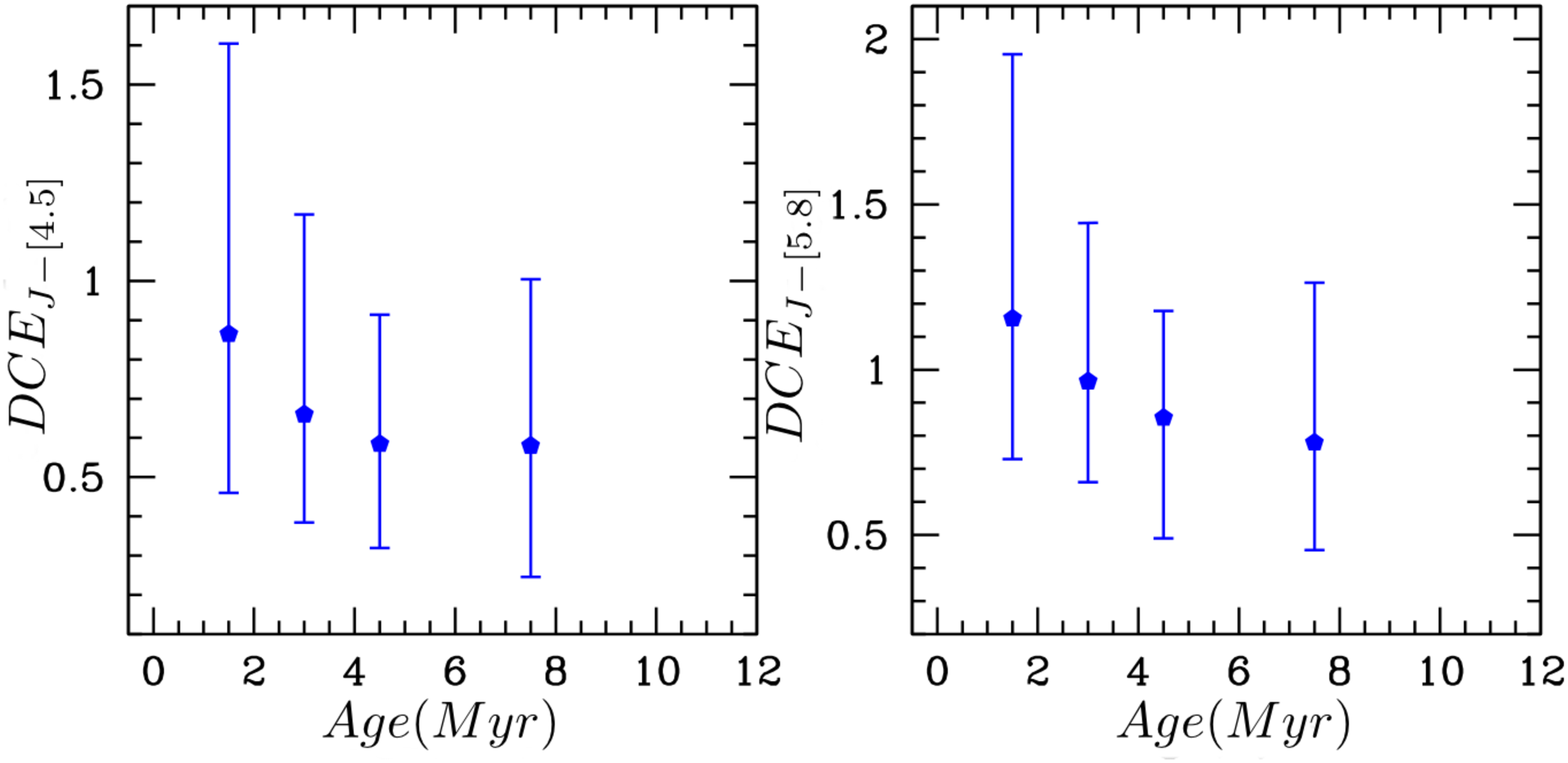}
\includegraphics[width=1\linewidth]{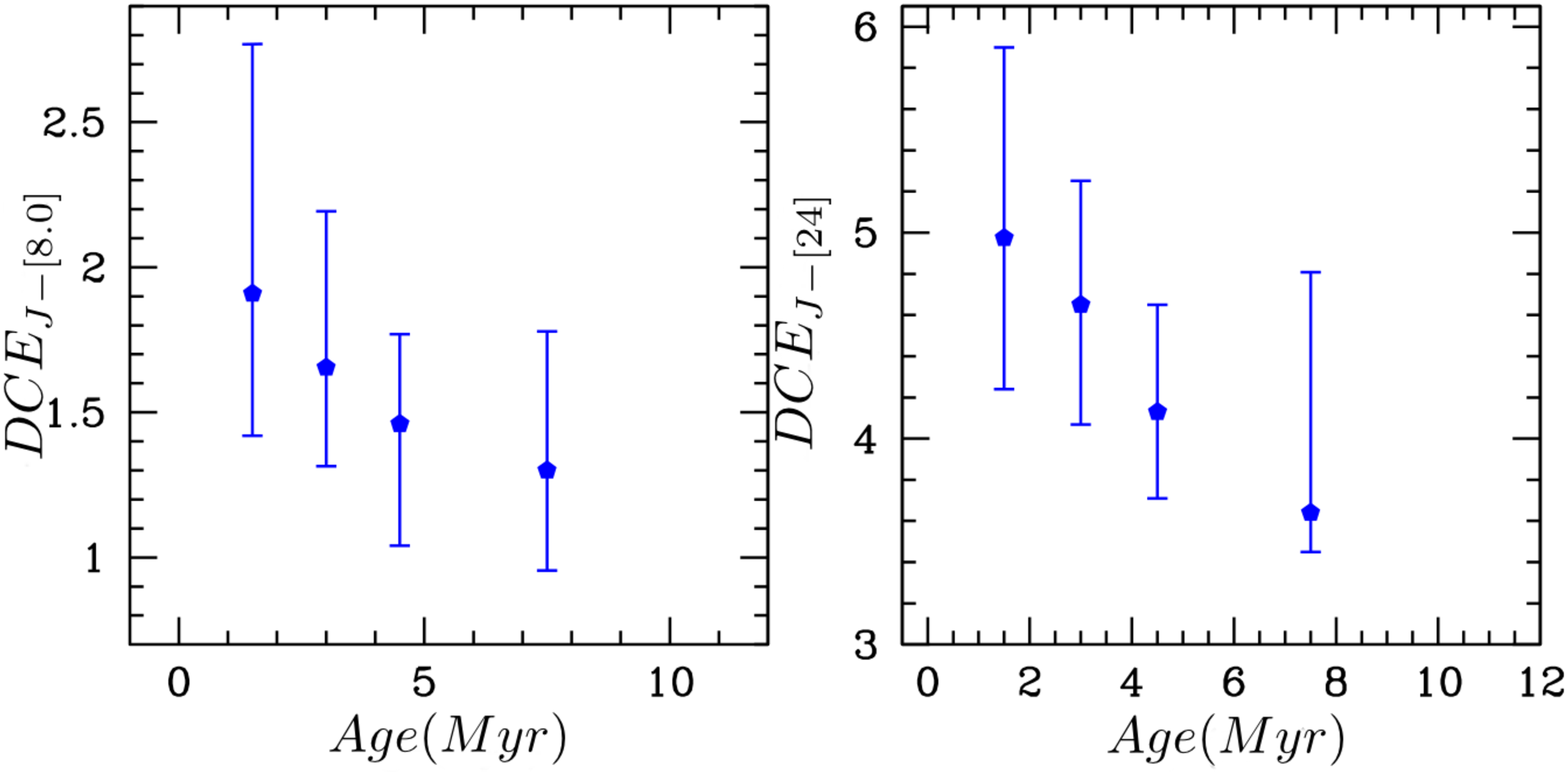}
\caption{\small Evolution of the median disk excess for: $a) \, J-[4.5]$ (upper left), $b) \, J-[5.8]$ (upper right), $c) \, J-[8.0]$ (lower left) and $d) \, J-[24]$ (lower right). The blue pentagons correspond to the median of each distribution for each age bin and the bars are the quartiles, which indicate the range where 50\% of the stars with excess are located.}
\label{fig_all_DCEs_vs_age_v2}
\end{figure}

For each age bin we build histograms for the observed DCEs in the IRAC and MIPS 24 bands, which are shown in Figure \ref{fig_all_DCEs_all_bins}, where we notice that the ``tails'' located at the right of the distributions (e.g. DCE$_{J-[4.5]}\gtrsim 2$) in the youngest bin, disappear as the age increases, while the position of the peak of the distributions remains approximately in the same position, with the exception of the color DCE$_{J-[24]}$, in which the peak moves towards smaller values with age. 
Figure \ref{fig_all_DCEs_vs_age_v2} shows
the median of 
the DCE distributions as a function of age; 
the quartiles, shown as error bars,  
indicate the range where 50\% of the disk-bearing stars are located. We performed a two-sample KS-test on the different DCE distributions for all the age bins and found the $p$-values shown in Table \ref{tab_KStest}. In general, these values indicate different DCE distributions between bin 1 and bin 2 ($p$-value $<$6\%) which 
suggest
significant dust evolution between these two bins. In contrast, the KS-test 
indicates no difference
between 
the DCE distributions
in 
bin 3 and bin 4 ($p$-value$>$90\%) 
suggesting
little
or no dust evolution between these older bins.

Although the distance to each star is not used to calculate the observed DCEs, an additional constrain to add to our observational sample 
would be to only include stars with GAIA DR2 parallaxes with relative errors $<20\%$ and with coherent parallaxes and proper motions, that is, within $3\times \sigma$ from the median value. 
As a test, we calculated the DCE distributions including
only disk-bearing stars with coherent parallax and proper motions in each group (restricted sample), and we obtained the same evolutionary trends, but with larger uncertainties, given by a larger range between the second and third quartiles.
Given this result and
since our goal is to make a statistical study with good significance,
we will work with the initial sample without further constrains. 
In any case, the stars in the our observational sample have previous membership confirmation for each stellar group.

\begin{deluxetable}{ccccc}
	\tablecaption{KS-test $p$-values
	\label{tab_KStest}}
\tabletypesize{\scriptsize}
\tablehead{
	\colhead{bins} & 
	\colhead{$DCE_{J-[4.5]}$}  &
	\colhead{$DCE_{J-[5.8]}$}  & 
	\colhead{$DCE_{J-[8.0]}$}  & 
	\colhead{$DCE_{J-[24]}$}  
}
\startdata
1-2  &0.04  &0.05 &0.06 &0.06 \\
2-3  &0.45  &0.23 &0.43 &0.17 \\
3-4  &0.99  &0.99 &0.99 &0.91 \\
\enddata
\end{deluxetable}

\subsection{Gas evolution: mass accretion rates}
\label{subsec_gas_evolution}

The mass accretion rate $\dot{M}$ 
measures the amount of disk material that falls onto the star per unit time and therefore it is an indicator of the
amount of gas in the disk.
Different studies have shown that 
the mass accretion rate
decreases with age \citep[cf.][]{Hartmann_2016},
which is expected from viscous evolution theories, since at the early stages the disk is 
fairly massive
and some mechanism 
is responsible for 
the redistribution of angular momentum, allowing a fraction of the disk material to fall onto the star. Another fraction of material moves to outer radii, since angular momentum is conserved.
Extensive determinations of mass
accretion rates in different star forming regions have been carried out
\citep[][and references therein]{Hartmann_2016},
but they have used different methods to measure the rates.
In here, we use our 
large sample of stellar
groups at different
ages to re-determine
the evolution of 
$\dot{M}$ in a homogeneous way. 

\subsubsection{Determination of
mass accretion rates}
\label{subsubsec_observed_mdots}

We determine $\dot{M}$ in our samples using the EW {H$\alpha$} estimated using the SpTClass tool \citep{Hernandez_2007b, Hernandez_2014, Hernandez_2017}. This estimation can be a good proxy to study the global evolution of the gas component of protoplanetary disks in young stellar objects \citep{Ingleby_2011, Alcala_2014, Mauco_2016, Fairlamb_2017}. As discussed in \S \ref{sec_sample}, the sample used to determine $\dot{M}$ includes some TTSs with measurements of EW {H$\alpha$}
without
IRAC photometry, and thus it is different from the sample analyzed in \S \ref{subsubsec_results_DCE_dist}. Additionally, the stars in this sample have GAIA DR2 parallaxes, with uncertainties below 20\%. Following the classification scheme from \citet{Briceno_2019}, we determine $\dot{M}$  for Classical TTSs (CTTSs) and for TTSs with EW {H$\alpha$} intermediate between that of a CTTS and
WTTSs (CWTTSs). Thus, we have included the following samples of CTTSs and CWTTSs (hereafter accretors) which have stellar masses estimated following \S\ref{sec_av}.\\ 

- \textit{Bin 1}: We have included 228 accretors located in the ONC (\S\ref{sec_onc}) and 16 accretors located in Taurus (\S\ref{sec_taurus}). Out of 244 accretors, 163 and 14 stars also belong to the sample used to study the dust in \S \ref{subsubsec_results_DCE_dist}, in the ONC and in Taurus, respectively.\\

- \textit{Bin 2}: We have included 40 accretors located in the $\sigma$ Ori cluster (\S\ref{sec_sori}). Out of 40 accretors, 36 stars also belong to the sample used to study the dust in \S \ref{subsubsec_results_DCE_dist}.\\

- \textit{Bin 3}: We have included 59 accretors located in Orion OB1b (\S\ref{sec_oriOB1b}). Out of 59 accretors, 19 stars also belong to the sample used to study the dust in \S \ref{subsubsec_results_DCE_dist}.\\ 

- \textit{Bin 4}: We have included 77 accretors located in Orion OB1a. (\S\ref{sec_oriOB1a}). Out of 77 accretors, 9 stars also belong to the sample used to study the dust in \S \ref{subsubsec_results_DCE_dist}.\\

\begin{deluxetable*}{lccccc}
	\tablecaption{Sample for mass accretion rates
	\label{tab_mdot}}
\tabletypesize{\scriptsize}
\tablehead{
	\colhead{Group} & 
	\colhead{Age} &   
	\colhead{Stars with} & 
	\colhead{Median $\log (\dot{M}$) }  &
	\colhead{$\sigma$}  &
	\colhead{Age}\\
	\colhead{} & 
	\colhead{Myr} &  
	\colhead{$\dot{M}$} & 
	\colhead{$M_{\odot}/yr$ }  &
	\colhead{}  &
	\colhead{bin} 
}
\startdata
ONC+Taurus & 1-3 & 244 & -8.49 &  0.50 & 1\\ 
$\sigma$ Ori & 3 & 40 & -8.60 & 0.32 & 2\\ 
Ori OB1b & 5 & 59 & -9.19 & 0.43 & 3\\
Ori OB1a & 9-14 & 77 & -9.29 & 0.50 & 4\\
\enddata
\end{deluxetable*}

The $\dot{M}$ was estimated from the {H$\alpha$} line luminosity ($L_{\text{H}\alpha}$)
using 
$L_{\text{H}\alpha}$=$4\pi\, d^2\times (\text{EW} \, \text{H}\alpha) \times F_{cont}$, where $d$ is the distance calculated as
the inverse of the GAIA DR2 parallaxes. The continuum flux at 0.6563 {\micron} ($F_{cont}$) was estimated
as follows. 
We calculated the
$\text{V}$ and $\text{I}_{\text{c}}$ magnitudes
of each object
from the
intrinsic $[\text{V}-\text{I}_{\text{c}}]$ colors from Table 6 of \citet{PM_2013} and
the $J$ magnitude corrected by extinction.
The  
V and $\text{I}_{\text{c}}$ magnitudes were converted to 
fluxes using the respective zero points for each filter,
and $F_{cont}$ was obtained by interpolating between these fluxes at 0.6563 {\micron}.

The accretion luminosity, $L_{\text{acc}}$, was computed using equation (10) from \citet{Ingleby_2013}

\begin{equation}
\log(L_{\text{acc}})=1.0(\pm 0.2)\log(L_{\text{H}\alpha})+1.3(\pm 0.7),
\label{eq_loglacc}
\end{equation}

\noindent
and 
the mass accretion rate for each star was then obtained from the accretion luminosity using the relation $L_{\rm acc} = {G M_{\ast} \dot{M}/R_{*}}$, where the derived stellar mass and
radius estimated from the effective temperature and the stellar luminosity (\S\ref{sec_av}) are used. Table \ref{tab_mdot} shows the groups for which the mass accretion rates were determined, their age, the median of $\log (\dot{M})$ and the standard deviation $\sigma$, including the stars with and without IRAC photometry. In Table \ref{megatable} we show all the stars with $\dot{M}$ shown in Table \ref{tab_mdot}, as well as the stars used in the determination of the different DCEs (\S \ref{subsubsec_results_DCE_dist}) without EW {H$\alpha$}. The errors associated with the different DCEs in Table \ref{megatable} are underestimated, since they do not include the uncertainties from the determination of spectral types, intrinsic colors or extinctions. We cannot include the propagated uncertainties, since most of the references from which the spectral types of effective temperatures were taken do not include such errors in the published data. These underestimated errors do not affect our results since we do not use them in any way as a weight to determine the disk parameters.

\subsubsection{Results: the evolution of the mass accretion rate}
\label{subsubsec_ev_mass_acc_rate}

\begin{figure}[h!]
\centering
\includegraphics[width=1\linewidth]{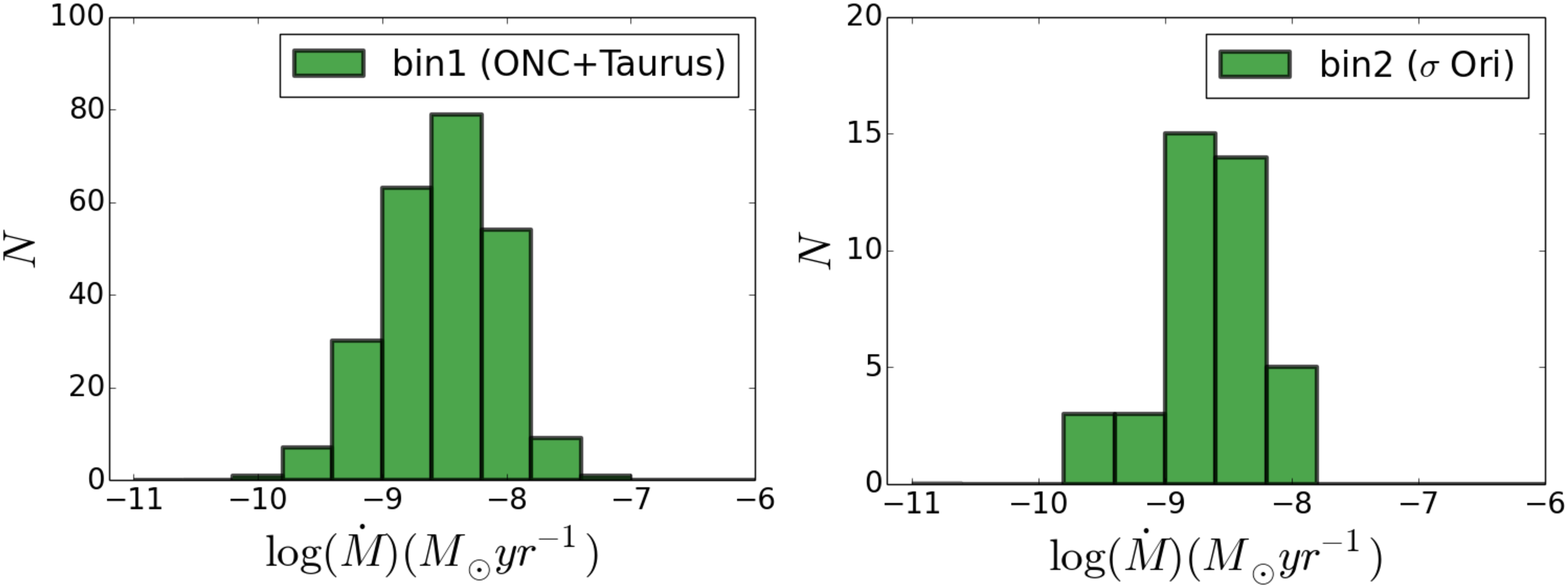}
\includegraphics[width=1\linewidth]{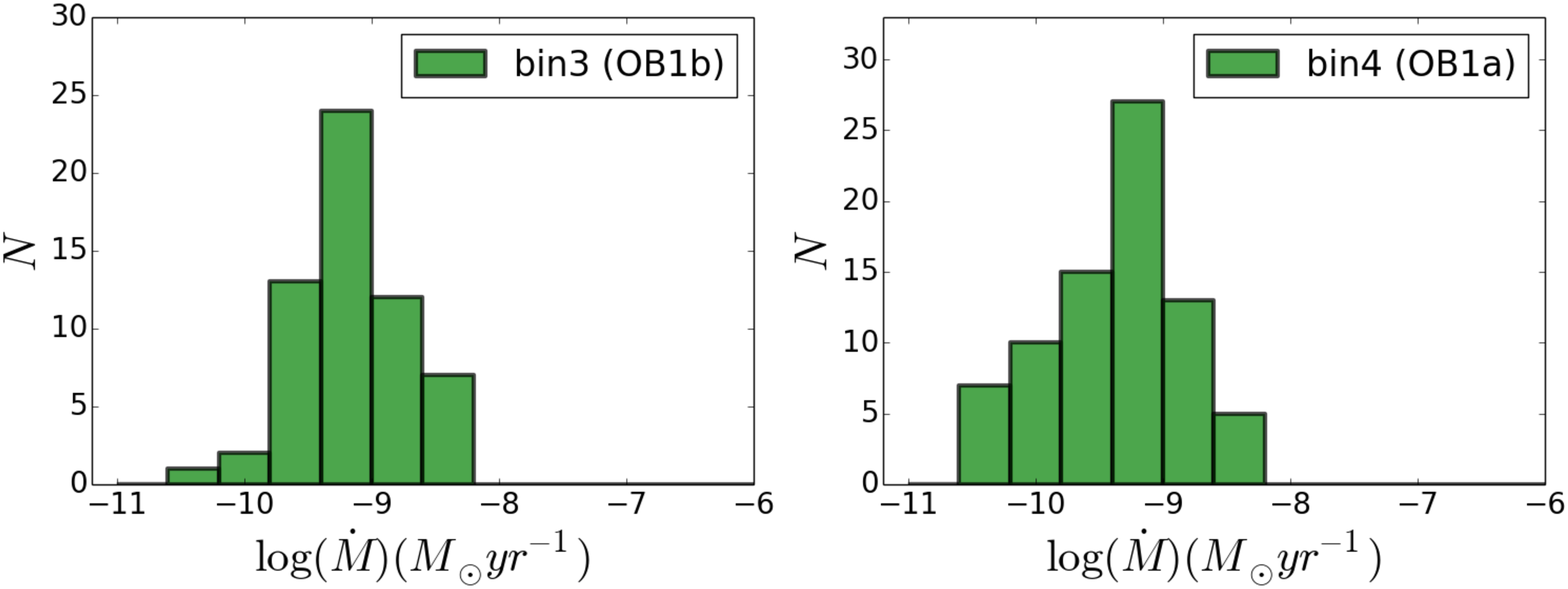}
\caption{\small Mass accretion rates distributions calculated in this work, for the different age bins.}
\label{fig_observed_mdots_2}
\end{figure}

\begin{figure}[h!]
\centering
\includegraphics[width=0.9\linewidth]{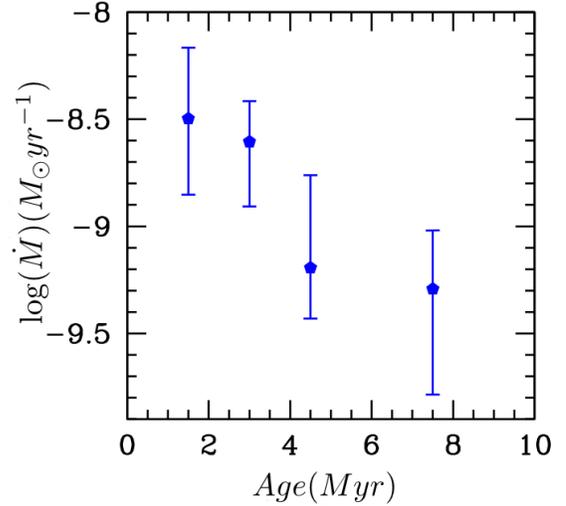}
\caption{\small Median of the mass accretion rates for the different age bins. The blue bars are the quartiles, which show where 50\% of the data of each age bin lie.}
\label{fig_median_logmdot_vs_age}
\end{figure}

In Figure \ref{fig_observed_mdots_2} we show the distribution of mass accretion rates of each age bin. It can be seen that the distributions, and their peaks, move towards lower values as the age of the group increases. Figure \ref{fig_observed_mdots_2} confirms what has been previously discussed and found in other works  \citep[e.g.][]{Hartmann_2016}. The mass accretion rates in this work were determined self-consistently using the same method for each stellar group. This is important because our results are not affected by any bias that might be introduced  when different methods are used. Figure \ref{fig_median_logmdot_vs_age} shows the median of the mass accretion rates vs the age bin. The bars are the quartiles, which show where 50\% of the data of each age bin lie. There is a clear trend of decreasing 
accretion rates with age, as
previously found.
The median  seems to fall more rapidly during the first 5 Myrs, and then the decrease flattens out.
Similar behavior is observed  in the evolution of the medians of the different DCEs (see Figure \ref{fig_all_DCEs_vs_age_v2}).

\section{Interpretation of observed evolution}
\label{sec_Interpretation of observed evolution}

\subsection{Evolutionary models}
\label{subsec_evolutionay_models}

\subsubsection{Mass accretion rates in viscously evolving disks}
\label{subsubsec_description_models}

We use the similarity solution of \citet{LP_1974} of a disk radial viscous evolution 
for the case of a power-law radial
dependence of the viscosity coefficient. In particular, for 
a viscosity that increases linearly with radius, $\nu \propto R$,
the surface density distribution as a function of time
$t$ can be written as

\begin{equation}
\Sigma(R,t)=\frac{M_{d}(0)}{2 \pi R_{1}^{2}}\frac{1}{r\mathbb{T}^{3/2}}\text{exp}\left[-\frac{r}{\mathbb{T}}\right],
\label{eq_sigma_gamma1}
\end{equation}

\noindent
the mass accretion rate at $R = 0$  as 
\begin{equation}
\dot M(0,t)=\frac{M_{d}(0)}{2 t_s}\frac{1}{\mathbb{T}^{3/2}},
\label{eq_mdot0}
\end{equation}

\noindent
and the disk mass as

\begin{equation}
M_d(R,t)=\frac{M_{d}(0)}{\mathbb{T}^{1/2}}\left\{1-\text{exp}\left[-\frac{r}{\mathbb{T}}\right]\right\}
\label{eq_mdisk}
\end{equation}

\noindent
\citep{Hartmann_1998}. Here, 
$M_{d}(0)$ is the initial disk mass, $R_{1}$ is a characteristic radius, inside which 60\% of the original disk
mass is located, $r = R / R_1$,
and $\mathbb{T}=t/t_{s}+1$ is a non-dimensional time, where $t_{s} = R_1^2/(3\nu_1)$ is the characteristic viscous timescale at $R_1$, and $\nu_1$ is the viscosity at $R_1$.

In the 
\citet{Shakura_1973} prescription, the viscosity is
given by
$\nu=\alpha c_{s}H$, where $H=c_s/\Omega_K$ is the scale height,
$\Omega_{K}=(GM_{\ast}/R^{3})^{1/2}$ is the Keplerian
angular velocity, where $G$ is the gravitational constant and $M_{\ast}$ is the stellar mass,
$c_{s}=(kT/\mu m_{H})^{1/2}$ is
the sound speed, where $T$ is the temperature, $\mu=2.34$ is the mean molecular weight, $k$ is the Boltzmann
constant, $m_{H}$ is the hydrogen mass, 
and
$\alpha$ is a free parameter.

Using this prescription, the viscosity is
$\nu \propto \alpha \, T \, R^{3/2}$. For 
an irradiated disk, and $R\gg R_{\ast}$,  $T \propto R^{-1/2}$
in which case $\nu \propto R$ and the 
similarity solution
can be applied
\citep[cf.][]{Hartmann_1998}.
Following this work,
we assume that this viscosity law
holds at each radius and use
equation (\ref{eq_sigma_gamma1})
to describe the evolution of the disk surface
density distribution.

Using the $\alpha$ viscosity prescription and
equation (\ref{eq_mdot0}),
the mass accretion rate onto the star
can be written as
\begin{equation}
\dot{M}(0,t)=\frac{3 \alpha}{2 (GM_{\ast}R_{1})^{1/2}}
\,
\frac{kT_{1}}{\mu m_{H}}
\,
\frac{M_d(0)}{\mathbb{T}^{3/2}},
\label{eq_mdot_new}
\end{equation}
for a temperature profile $T=T_1(R/R_1)^{-1/2}$, such that the viscous time is
$ t_{s}=(1/3 \alpha)(GM_{\ast}R_{1})^{1/2}(\mu m_{H})/(k T_1)$. 
For a given initial disk mass $M_{d}(0)$ and
a characteristic radius $R_1$,
and for a star of mass $M_*$, equation (\ref{eq_mdot_new})
gives the predicted evolution of
the mass accretion rate onto the star.
Since for the stellar groups discussed in \S \ref{sec_sample}, $t >> t_s$, one obtains

\begin{equation}
\dot{M}(0,t)=\frac{\left(GM_{\ast}R_{1}\right)^{1/4}}{2 \left(3 \alpha \right)^{1/2}} \left(\frac{\mu m_{H}}{kT_{1}} \right)^{1/2} 
\frac{M_d(0)}{{t}^{3/2}}.
\label{eq_mdot_new_1}
\end{equation}

We adopt the approximation
$T_1 = T_{d2} (100 \, {\rm au} / R_1)^{1/2}$,
with $T_{d2} = 10$K, a typical temperature
at 100 AU \citep[cf.][]{Dalessio_1998,Dalessio_2001}. Then, the mass accretion rate can be written as 

\begin{eqnarray}
\frac{\dot{M}(0,t)}{\msun \, {\rm yr^{-1}}}& = & 1.39 \times 10^{-8} \left(\frac{\alpha}{0.01}\right)^{-1/2} \left(\frac{M_{\ast}}{\msun} \right)^{1/4} \left(\frac{R_1}{\rm AU}\right)^{1/2} \nonumber \\
& & \times \left(\frac{T_{d2}}{10 K} \right)^{-1/2} \left(\frac{M_d(0)}{\msun}\right) \left(\frac{t}{\rm Myr}\right)^{-3/2}.
\label{eq_mdot_norm}
\end{eqnarray}

In the fiducial model of \citet{Hartmann_1998}, the parameters
$M_{d}(0)$ and $R_{1}$ were set to 0.1 $\msun$ and 10 au,
respectively, and calculations were
shown for $M_* = 0.5 \, \msun$. In here, we will use the values 
for $\alpha$, $M_{d}(0)$, $R_{1}$, and 
 $M_*$, that are consistent with our observational samples and the observed decay of mass accretion rate with age
(cf. Figure
\ref{fig_median_logmdot_vs_age}).

We adopt a stellar mass 
$M_* = 0.30 \, \msun$, which corresponds to the mean stellar mass from our sample (see \S  \ref{subsec_separation}),
as representative of the evolution
of the mean color excesses. We fit the observed decay of the mass accretion rate with age
(Fig. \ref{fig_median_logmdot_vs_age}) using equation (\ref{eq_mdot_norm}) and find the triads of values ($R_1, M_{d}(0), \alpha$) that provide the 
best $\chi^2$ fit to the data.

\begin{figure}[h!]
\centering
\includegraphics[width=1\linewidth]{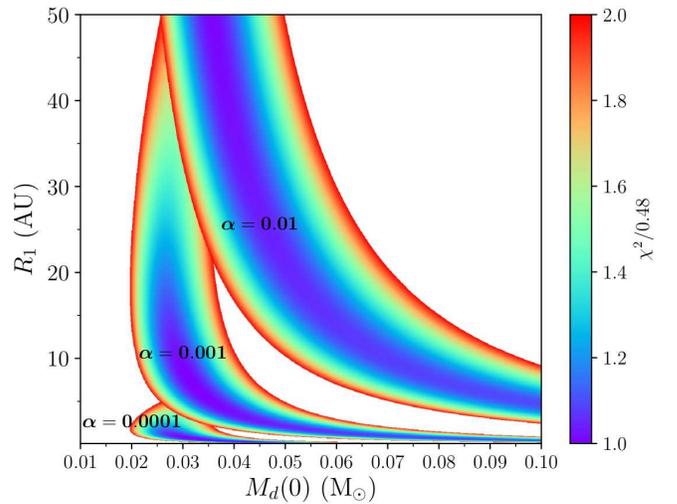}
\caption{\small
Normalized $\chi^2/0.48$ for the
fits to the observed decay of
mass accretion rate with age in Figure
\ref{fig_median_logmdot_vs_age} from different
combinations of
the evolutionary parameters
$R_1$ and $M_d(0)$, and
the viscosity parameter
$\alpha$. Each color map region corresponds
to a different value of $\alpha$,
indicated in the label. The values are normalized to the global minimum as indicated in the scale.
}
\label{fig:chisquared}
\end{figure}

Figure 
\ref{fig:chisquared} shows
the goodness of the fits,
as indicated by the value of
$\chi^2$, for 
a range of values of $R_1$ and $M_d(0)$, 
and for three values of $\alpha$: 0.01, 0.001, and 0.0001.

The lowest value of $M_{d}(0)$ among the best fitting models for $\alpha$=0.001 is about 0.03 $M_{\odot}$, or about twice the minimum solar nebula (MMSN). The mean value of the disk mass for disks in the 2 Myr old Taurus association is 0.005 $M_{\odot}$
\citep{Andrews_2005}, so in principle the mean initial disk mass could be lower than 0.03 $M_{\odot}$. However, \citet{Andrews_2005} note that the formation of giant planets requires disk masses of a few MMSN.
In addition, models with
$\alpha \lesssim 0.001$ predict very low values
of $R_1$, less than 5 au. In contrast, 
a large ALMA disks' survey in Orion
finds   
that Class 0 and Class I objects have mean dust disk radii of $\sim 45$ au and $\sim 37$ au, respectively \citep{Tobin_2020}. These objects are representative of the initial stages of viscous evolution, which argues against very low values of $R_1$. 

\begin{figure}[h!]
\centering
\includegraphics[width=1.0\linewidth]{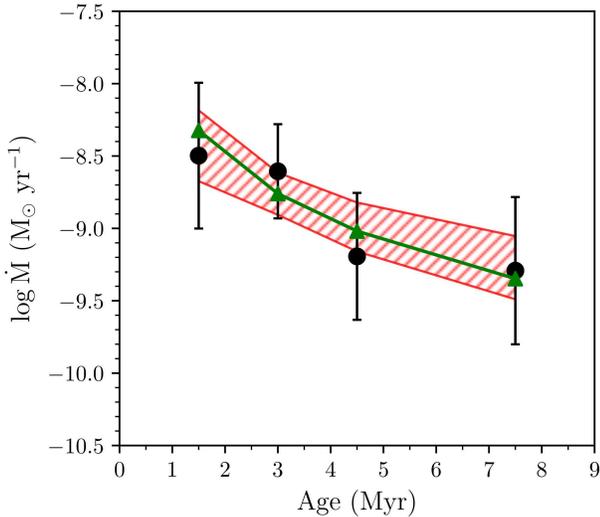}
\caption{\small Median mass accretion rates as a function of age (black points), where the error bars correspond to the standard deviation of the distributions at each age. The solid green line shows the theoretical mass accretion
rate for $M_d(0)=0.07 \, M_{\odot}$,  $R_1=10$ au,  and $\alpha$=0.01. The red region corresponds  to  models  with $\chi^2/0.48 \le 2 $  in Figure~\ref{fig:chisquared}.}
\label{fig_mdot_model}
\end{figure}

For $\alpha=0.01$, Figure \ref{fig:chisquared} shows that values of $R_{1}$ ranging from $\sim 5$ au to radii even larger than 50 au, with $0.03 \, M_{\odot} \lesssim M_{d}(0) \lesssim 0.1 \, M_{\odot}$, give reasonable fits to the observed data (cf. Figure \ref{fig_mdot_model}). The observed mean values of the dust disks' radii for Class 0 and Class I objects from \citet{Tobin_2020}, suggest that the characteristic radius $R_{1}$ is larger than the values inferred from the analysis for $\alpha=0.001$ and 0.0001. To be consistent with these considerations, we adopt models with $\alpha = 0.01$.
 
Figure \ref{fig_mdot_model} shows the median values of the mass accretion rate distributions at each age bin (also shown in Figure \ref{fig_median_logmdot_vs_age} and Table \ref{tab_mdot}). 
The standard deviation of the distributions of mass accretion rates (shown with the error bars) was calculated using the median absolute deviation (MAD) of the observational data, 
assuming Gaussian distributions,
with the relation $\sigma = 1.4826 \, \text{MAD}$. 
The red shaded area in Figure \ref{fig_mdot_model} corresponds to the 
mass accretion rates predicted
by triads  of values ($R_1, M_{d}(0), \alpha$) which have normalized $\chi^2/0.48 \le 2$
in Figure \ref{fig:chisquared},
and 
the green solid line shows the theoretical mass accretion
rate for one of these triads, 
 $R_1=10$ au \citep[e.g.][]{Hartmann_1998}, $M_d(0)=0.07 \, M_{\odot}$, and $\alpha$=0.01. 
This analysis suggests that the observed decay of $\dot{M}$ with age is consistent with viscous evolution of the gas in the disk, for these evolutionary models.

\subsubsection{D'Alessio Models}
\label{subsubsec_dalesio_model}

We aim to calculate the emission of disks
for which the radial distribution of surface density
is given by equation (\ref{eq_sigma_gamma1}) at each one of the age bins
for which we have obtained color excesses.
To do so, we need to calculate the vertical structure
of the disk, namely, we need to solve
the disk structure equations to obtain
the 
radial and vertical
distribution of the temperature and
volumetric density which result in the
given $\Sigma(R,t)$. For this purpose,
we use the irradiated disk models developed by D'Alessio (1996), \citet{Dalessio_1998, Dalessio_1999, Dalessio_2001, Dalessio_2006}, adopting several approximations to adapt these models to the \citet{LP_1974} models.

The D'Alessio models assume that the disk is
steady, geometrically thin and axially symmetric. The dust and gas are assumed
to be thermally coupled.
The disk is heated by absorption of stellar and accretion shock radiation as well as viscous dissipation, and it is cooled by dust emission. The viscosity
is given by the \citet{Shakura_1973} prescription.
 
As discussed,
with this prescription
the viscosity only increases linearly with
radius, as assumed in the \citet{LP_1974}
similarity solution, if the temperature depends
on radius as $T \propto 1/R^{1/2}$ \citep{Hartmann_1998},
which only happens approximately in 
more realistic physical situations \citep[cf. ][]{Dalessio_2001, Dalessio_2006}. Nonetheless,
we will adopt the $\alpha$ prescription to represent
the viscosity, with the sound speed evaluated at
the midplane temperature.

Another issue is that in the D'Alessio models
the surface density is that of a steady disk
and lacks the exponential fall-off at large
radii (cf. equation \ref{eq_sigma_gamma1} ). Following \citet{Qi_2011},
we modified the $\alpha$ parameter so that $\alpha=\alpha_{0}e^{R/R_{t}}$, where $R_{t}$ is the critical radius outside which the exponential fall-off becomes important
and $\alpha_{0}$ is the value of $\alpha$ at $R << R_t$. The critical radius
is given
by $R_t = R_1 \mathbb{T}$ (equation \ref{eq_sigma_gamma1}). For
$t >> t_s$, this is

\begin{eqnarray}
R_t & = &1.37 \times 10^2 {\rm AU}  \left(\frac{T_{d2}}{10 \, {\rm K}}\right)  \left( \frac{M_*}{0.30 \, M_\odot}\right)^{-1/2} \nonumber \\
&  & \left(\frac{\alpha_1}{0.01}\right)  \left( \frac{t}{{\rm Myr}}\right).
\label{eq:rt}
\end{eqnarray}

As stated in \citet{Qi_2011},  this modification adds an exponential taper to the radial distribution of the surface density, and does not reproduce the predicted behavior
of the mass flow at large radii, but the effect on the
observables is negligible, since irradiation dominates
over viscous heating at 
these large radii.

The main input parameters for the D'Alessio
models are the stellar mass, radius, and effective
temperature, 
the mass accretion rate and viscosity
parameter,
the dust composition, size distribution,
and spatial distribution in the disk.
To be consistent with the surface density
distribution resultant from viscous 
evolution calculations at a given
age bin
(equation \ref{eq_sigma_gamma1}), the model
uses as further input
the transition radius $R_t$,
equation (\ref{eq:rt}), with $\alpha = 0.01$.

In addition,
we use the evolutionary models of \citet{Siess_2000} 
for a star of mass  0.30 $ M_{\odot}$ 
(\S \ref{subsec_separation} and \S \ref{subsubsec_description_models}) to obtain the
stellar
luminosity and effective temperature at each 
age bin (see Table \ref{tab_models}).

In the models, the disks are irradiated by 
the star and the accretion shocks on the stellar
surface.
The shocks are formed when disk material falls onto the star, creating hot spots on the  surface, with a characteristic temperature of $\sim 8000 \, \text{K}$
\citep{Hartmann_2016}. 
The total luminosity irradiating the disk is
then
$L_{\text t} = L_* + L_{\text{shock}}$,
where we adopt
$L_{\text{shock}} = L_{\text{acc}}=GM_{\ast}\dot{M}/R_{\ast}$.

The dust species used in the models are silicates and graphite, with abundances typical of the ISM 
\citep{Draine_1984}. We assume a power law size distribution for
the grain sizes, with exponent 3.5,
between a minimum grain radius $a_{\text{min}}=0.005 \, \mu m$ and 
a maximum radius $a_{\text{max}}$.
Dust settling  
is included in the models by assuming two populations of dust grains with different values of $a_{\text{max}}$. The  
disk upper layers contain small grains, while 
the large grains are concentrated towards
the midplane 
\citep{Dalessio_2006}. We use $a_{\text{max}}=0.25 \, \mu m$ for the upper layers, and $a_{\text{max}}=1 \, \text{mm}$ in the midplane \citep{Dalessio_2006}. Both populations have different dust-to-gas mass ratios, $\zeta$. Because the midplane region contains the mass of the dust grains that have settled  from the upper layers, the corresponding $\zeta_{\text{big}}$ is larger than the standard value $\zeta_{\text{std}}$=0.01 \citep{Draine_1984}. Correspondingly, in the surface layers $\zeta_{\text{small}}$ is smaller than the standard value. 
The large grain population extends to a height 
$z_{\text{big}}(r)$, and
we choose $z_{\text{big}}(r) = H$ in this work,
where $H$ is the scale height
at the local temperature. 
The degree of dust settling is quantified using the parameter
$\epsilon=\zeta_{\text{small}}/\zeta_{\text{std}}$ \citep{Dalessio_2006}. 
The values for $\zeta_{\text{big}}/\zeta_{\text{std}}$ are computed
using equation (A5) from \citet{Dalessio_2006}.

The disk structure is calculated self-consistently by solving the vertical structure equations in an iterative scheme \citep{Dalessio_1998}.
The radiative transfer is done 
using mean opacities calculated
in two wavelength regions, 
the ``stellar'' and the ``local'' temperature
regions \citep{Calvet_1991, Dalessio_1998}. The mean opacities are
computed from the
monochromatic opacities
of each dust
species. The ``stellar'' mean opacities are the average of opacities calculated at the stellar
and shock temperatures, weighted by the corresponding luminosity. 

Once the disk structure is computed, the 
emission from the disk for a given inclination
angle $i$
is calculated with a ray-by-ray solution of the
transfer equation, including thermal
emission from the dust and a single
scattering of stellar radiation
\citep{Dalessio_1998,Dalessio_2006}.

The most important contributor to the
overall disk emission in the near-IR is the 
disk region at the dust destruction
radius \citep{Isella_2005, Dalessio_2005,Dullemond_2010}, which we name ``the wall''.
For radii smaller than the wall radius
the temperature reaches values above the 
silicates sublimation temperature ($\sim 1500$ K),
and 
we assume that dust 
does not exist. 
The wall is 
generally assumed to be
vertical and irradiated frontally by the central star, but 
several studies have shown that the
wall may instead be curved \citep{Isella_2005, Tannirkulam_2008, McClure_2013}.
This kind of shape for the wall is expected because the larger dust grains settle towards the disk midplane; 
since the micron-sized atmospheric grains are more efficient absorbing stellar radiation,
the larger grains can survive closer to the star. In addition, 
the sublimation temperature 
decreases with density \citep{Pollack_1994}.
\citet{McClure_2013} modeled the SED and near-IR spectra of four T Tauri stars in the Taurus region and found that a  
treatment of the wall that accounts for its curvature allows
for a much better fit
to the spectra than
a vertical wall.
Following \citet{McClure_2013} we    
mimic
the curvature of the wall as a two-layer wall. The lower layer contains 
larger ($1 \, \mu m$) hotter grains (1500 K),
while the upper layer contains sub-micron (0.25 $\mu m$), cooler (1000 K) grains. The radius of each layer, $R_{in_{1}}$ and $R_{in_{2}}$ for the lower and upper wall, respectively,
is set by the total luminosity, $L_{\rm t}$, and  the
respective
dust properties, using the relation $R_{in_{i}}\propto(L_{\ast}+L_{\text{acc}})^{1/2}(T_{w_{i}})^{-2}$, where $T_{w_{i}}$ is the sublimation temperature of the dust grains in the lower ($i=1$) and the upper ($i=2$) wall, respectively.
Both layers are frontally illuminated with a normal impinging angle. 
The calculation of the emission
from each layer 
follows the treatment of \citet{Dalessio_2005}.

The total SED is constructed by adding the
emission from the 
disk, the 
wall,
the star, and the accretion shock. We
assume the shock emits as a blackbody with
$T = 8000$ K and 
emitting
area $f 4 \pi R_*^2$. 
The filling factor $f$ is then calculated
by the condition that the shock luminosity
is the accretion luminosity.

\subsection{Comparison with observations}
\label{subsec_comparison_obs}

With gas evolution constrained
by the observations of mass accretion rates
onto the star at different ages (\S \ref{subsubsec_ev_mass_acc_rate}) , 
we aim to find the
dust properties that better describe
the observed evolution of
the disk color excesses
(\S \ref{subsubsec_results_DCE_dist}).
Specifically, we 
constructed grids of disk models at each age bin and calculated
the SEDs and color excesses for
each model in the grids (\S \ref{subsubsec_grid}). 
We used the models' color excesses
to
estimate the distributions
of the degree of dust settling
and of the inner wall height that
best fitted the observed distribution 
of color excesses at each age.

\subsubsection{Construction of the model grid}
\label{subsubsec_grid}

Using the  
methods described in Section \ref{subsubsec_dalesio_model}, we computed a grid of disk models for each age bin. Models were calculated using the following mass accretion rates: $2\times 10^{-8}, 1\times 10^{-8}, 5\times 10^{-9}, 1\times 10^{-9}$ and $5\times 10^{-10} M_{\odot}yr^{-1}$.
For simplicity, we only considered cases where the upper and the lower parts
of the two-layered wall at the
dust destruction
radius 
have the same ratio $z_{wall}/H=z_{wall_{1}}/H_{1}=z_{wall_{2}}/H_{2}$, where $H_{1}$ and $H_{2}$ are the gas scale heights $H_{1}\propto T_{w_{1}}^{1/2}R_{in_{1}}^{3/2}M_{\ast}^{-1/2}$ and $H_{2}\propto T_{w_{2}}^{1/2}R_{in_{2}}^{3/2}M_{\ast}^{-1/2}$ at the lower and upper wall, respectively.

\begin{figure}[h!]
\centering
\includegraphics[width=1\linewidth]{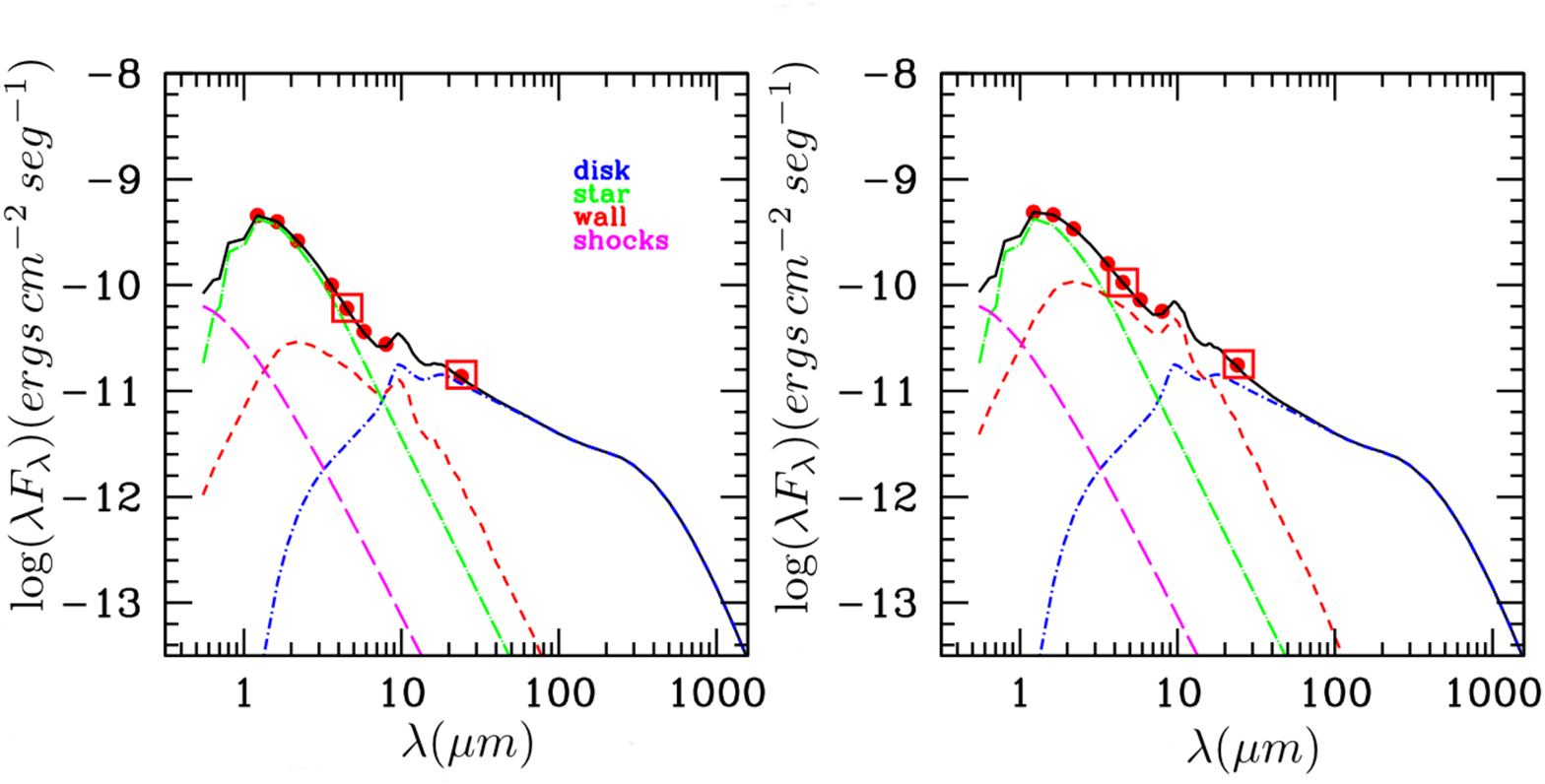}
\caption{\small SEDs of disk models from bin 1, with $\epsilon$=0.0001, $z_{wall}/H$=0.8 (left) and $z_{wall}/H$=3 (right), with $\dot{M}=1.0\times 10^ {-8} \, M_{\odot}/yr$. Each line represents a different component: the star (green dashed), the wall (red short-dashed), the accretion shocks (pink long-dashed), the disk (blue dot-dashed), and the total (black solid). The red points correspond to the 2MASS, IRAC, and MIPS 24 fluxes; the red squares indicate the flux at 4.5 $\mu m$ and at 24 $\mu m$. The inclination is $\cos(i)=0.5$.}  
\label{fig_SED}
\end{figure}

\begin{figure}[h!]
\centering
\includegraphics[width=1\linewidth]{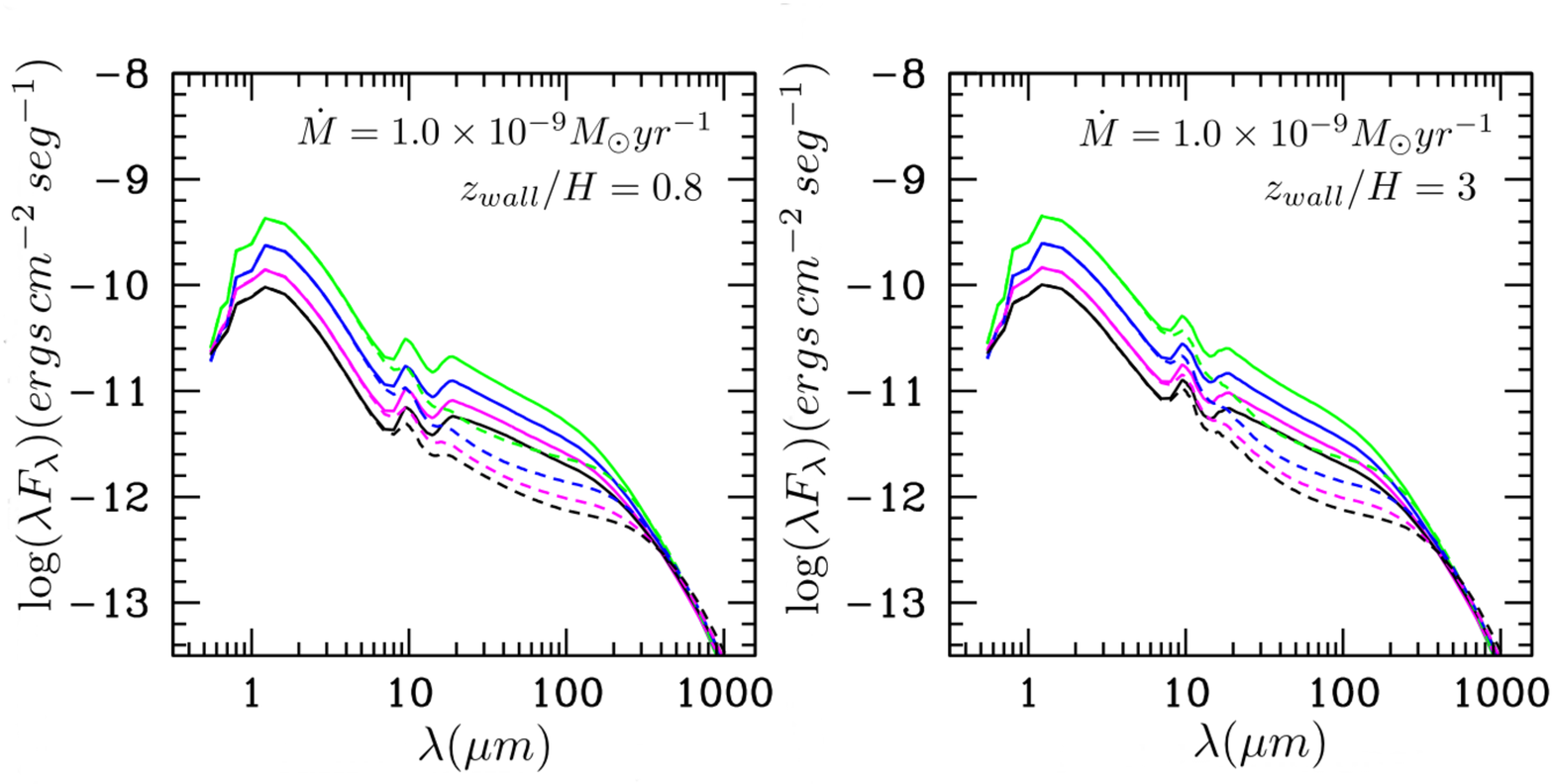}
\includegraphics[width=1\linewidth]{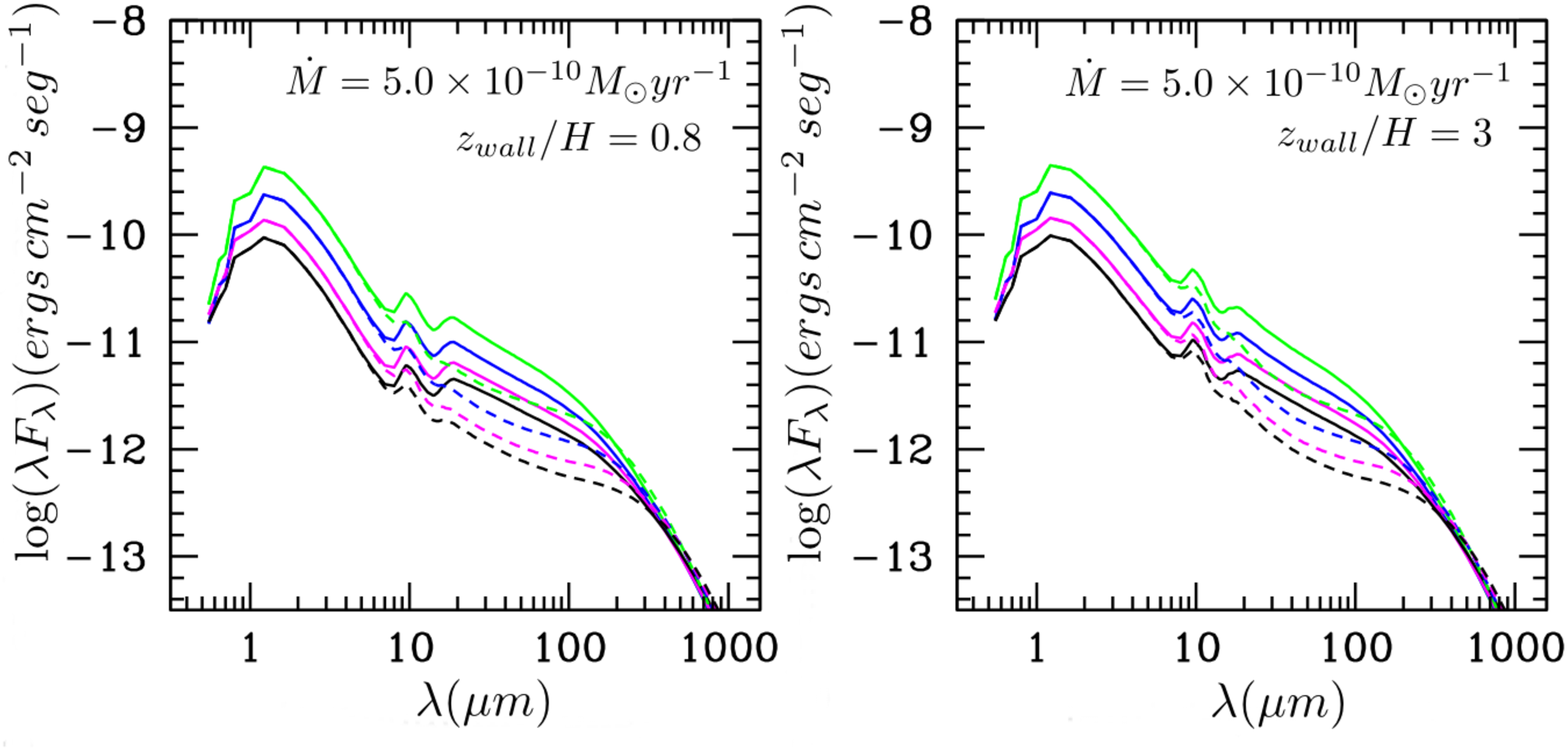}
\caption{\small SEDs of disk models for different age bins: bin 1 (green), bin 2 (blue), bin 3 (magenta), and bin 4 (black), with $\dot{M}=1.0\times 10^ {-9} \, M_{\odot}/yr$ (upper panels) and $\dot{M}=5.0\times 10^ {-10} \, M_{\odot}/yr$ (lower panels). Two values of $\epsilon$ are shown: 0.01 (solid lines) and 0.0001 (dashed lines). The left and right panels correspond to models with $z_{wall}/H$=0.8 and  $z_{wall}/H$=3, respectively. The inclination is $\cos(i)=0.5$.}  
\label{fig_SED_2}
\end{figure}

As mentioned before, we assume that the dust in the lower and upper wall sublimates at $T_{w_{1}}=1500$ K (with $a_{max}=1 \, \mu m$) and $T_{w_{2}}=1000$ K (with $a_{max}=0.25 \, \mu m$), respectively. We run models with values for 
$z_{wall}/H$ from 0.1 to 4, in steps of 0.8
. The total physical height of this two-layered wall is then
$Z_{wall}(au)=(z_{wall}/H) \, (H_1 +H_{2})$.
We also varied the settling parameter 
$\epsilon$ (from 1 to 0.0001), 
and cosine of the inclination
$\cos (i)$ (from 0 to 1). For each age bin we run 1350 disk models, thus the total number of models calculated, including the four bins, is 5400. Table \ref{tab_models} shows the parameters of the models, adopted in each age bin. Since we are not including nor modeling data beyond $24 \, \mu m $, the disk outer radius $R_{\text{disk}}$ is chosen arbitrarily, such that it is located beyond the critical radius $R_{t}$. We use these parameters to compute disk models and build their synthetic SEDs and DCEs (see Figures \ref{fig_SED} and \ref{fig_SED_2}).

\begin{deluxetable}{lcccc}
	\tablecaption{Parameter of the models
	\label{tab_models}}
\tabletypesize{\scriptsize}
\tablehead{
	\colhead{Age} &
	\colhead{$R_{\ast}$}  & 
	\colhead{$T_{\text{eff}}$} &  
	\colhead{$L_{\ast}$} & 
	\colhead{$R_{\text{disk}}$}\\
	\colhead{Myr} & 
	\colhead{$R_{\odot}$}  & 
	\colhead{$K$} &  
	\colhead{$L_{\odot}$} & 
	\colhead{au}  
}
\startdata
1.5 & 1.71     & 3360 & 0.38 & 230\\
2.5 & 1.25     & 3385 & 0.21 & 410\\
4.5 & 0.97     & 3416 & 0.13 & 660\\
7.5 & 0.79     & 3451 & 0.09 & 1080\\
\enddata
\end{deluxetable}

To compute the DCEs of the models in the IRAC and MIPS 24 bands
we calculate their  
fluxes and magnitudes in the 2MASS, IRAC and MIPS 24 bands,
using the corresponding 
transmission profiles of the filters. Using the $J$ band and the IRAC and MIPS 24 bands, we  
then built the corresponding synthetic colors. 
We also obtain the values for the photospheric magnitudes and photospheric colors of the models,
and calculate the
corresponding DCEs, by subtracting the models' photospheric colors from the colors obtained from the complete synthetic SED of the model.
We calculate the DCEs 
for the
colors $J$-[4.5], $J$-[5.8], $J$-[8.0] and $J$-[24].

Figure \ref{fig_SED} shows the SEDs of disk models with stellar parameters from bin 1, for $z_{wall}/H=$0.8 (left panel) and  $z_{wall}/H=$3 (right panel).
It can be seen  
that at 24 $\mu m$ 
the flux is dominated by the disk contribution.
Accordingly, the excess at
24 $\mu m$ can be used to estimate the degree of dust settling in the disk, quantified by the parameter $\epsilon$ (\S \ref{subsubsec_dalesio_model}),  
which measures the dust depletion in the disk upper layers \citep{Dalessio_2006}. In Figure \ref{fig_SED_2} we show the SEDs of disk models with $z_{wall}/H$=0.8 (left panels) and $z_{wall}/H$=3 (right panels), for bin 1 (green), bin 2 (blue), bin 3 (magenta), and bin 4 (black). In the upper and lower panels we show models with $\dot{M}=1.0\times 10^ {-9} \, M_{\odot}/yr$ and $\dot{M}=5.0\times 10^ {-10} \, M_{\odot}/yr$, respectively; additionally, two values of the parameter $\epsilon$ are indicated with different lines: 0.01 (solid lines) and 0.0001 (dashed lines).

\begin{figure}[h!]
\centering
\includegraphics[width=1\linewidth]{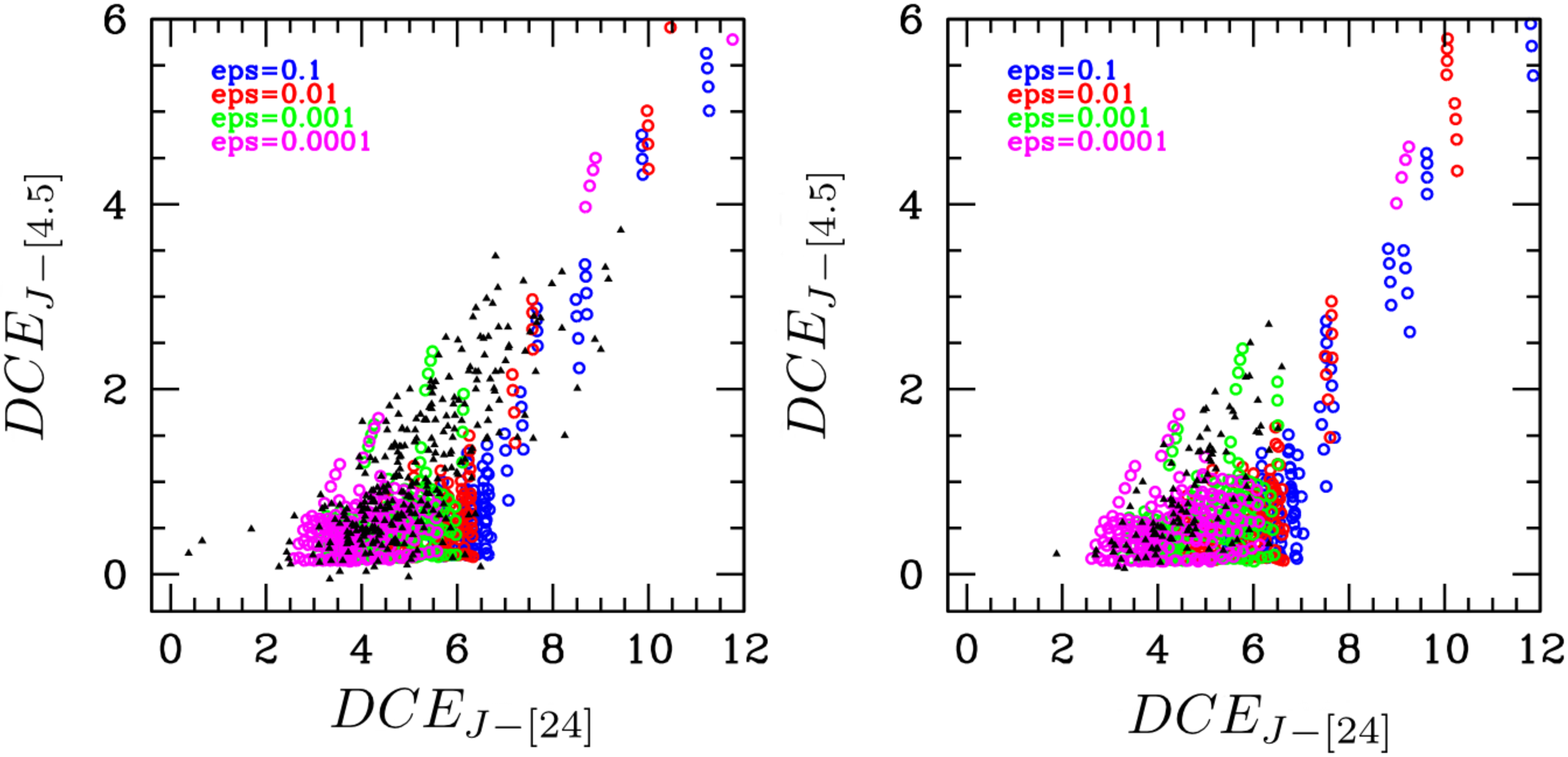}
\includegraphics[width=1\linewidth]{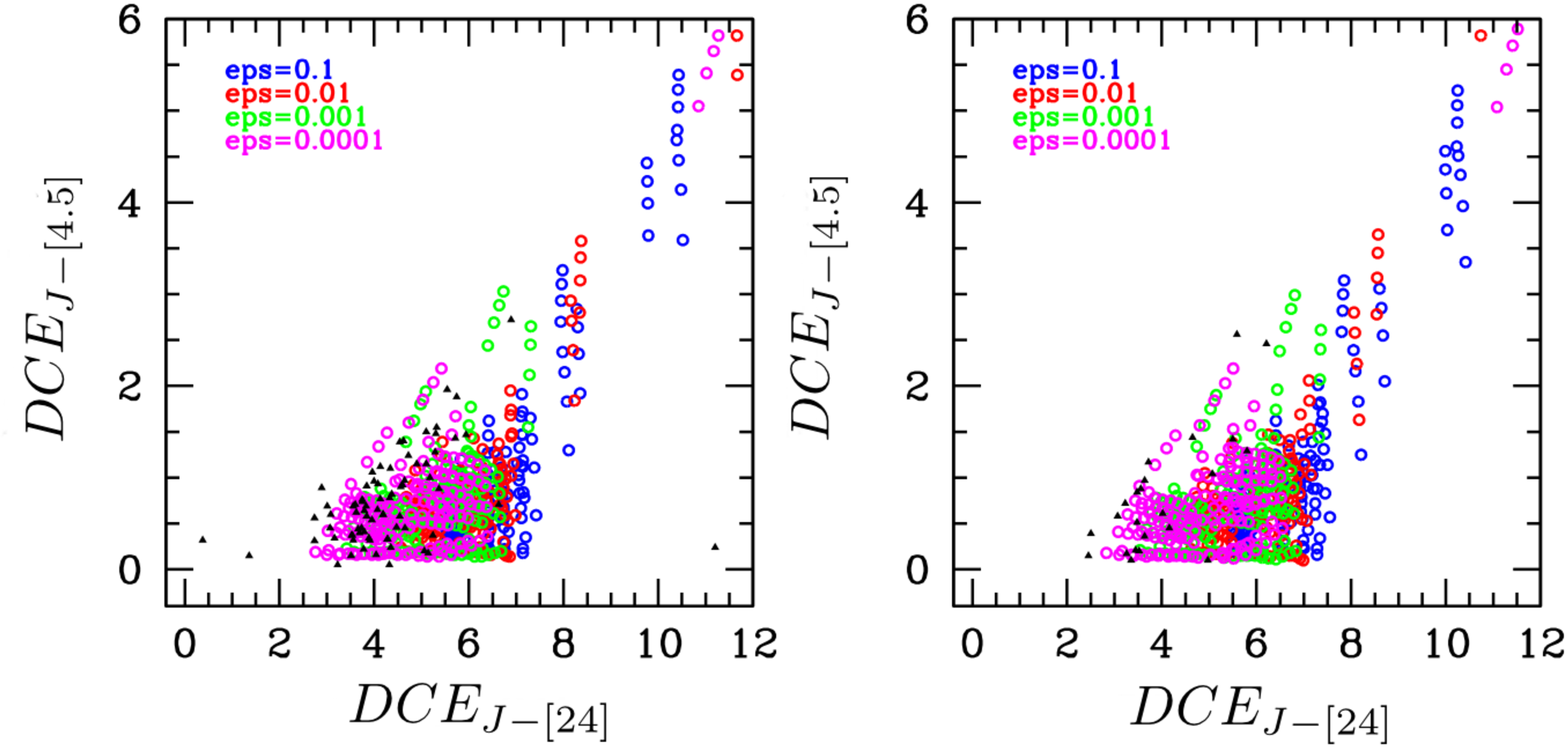}
\caption{\small Synthetic DCEs obtained from the grid of models for bin 1 (upper left), bin 2 (upper right), bin 3 (lower left) and bin 4 (lower right). Various $\epsilon$ are shown in different colors: 0.1 (blue), 0.01 (red), 0.001 (green) and 0.0001 (magenta). The observational sample in each age bin (black triangles) is shown for comparison.}  
\label{fig_DCEs_models}
\end{figure}

In Figure \ref{fig_DCEs_models} we plot the $DCE_{J-[4.5]}$ vs $DCE_{J-[24]}$ from the corresponding grid of disk models used in each age bin (empty circles), showing in different colors various values of $\epsilon$: 0.1 (blue), 0.01 (red), 0.001 (green) and 0.0001 (magenta). We also plot the observational sample in each bin (black solid triangles) for reference. Both the observational data and the models locate in the same region. The models with the largest $DCEs$ (e.g. $DCE_{J-[4.5]}\gtrsim 3.5$ and $DCE_{J-[24]}\gtrsim 8$) correspond to models of highly inclined disks, which are more difficult to detect and observe. Notice that, as the age of the bin increases, the observed disks move towards the smallest values of $\epsilon$, which is in agreement with the results we find using approximate Bayesian computation (see \S \ref{subsubsec_results_medians}).

\subsubsection{Spatial distribution of dust: forward modeling}
\label{subsubsec_statistical_app}

In the previous Section \ref{subsec_evolutionay_models} we used a model to define a sampling of disk DCEs for each age bin. Our observed data comprise distributions of DCEs as shown in Figure \ref{fig_all_DCEs_all_bins}. Our goal then is to take our grid of models and determine the underlying parameter values which would recreate the observed DCE histograms. We will use approximate Bayesian computation (ABC) to quantify the posterior probability distributions for each parameter.

ABC was developed as a way to apply Bayesian statistical methods when an analytical representation of the data likelihood function is replaced with simulation-based models  \citep{Turner_2001}. When using ABC, the forward model grid allows us to sample directly from the posterior by generating simulated datasets that match the observations. ABC requires a forward model, a summary statistic to use on the data and on the simulations, and a rule-set to accept or reject the choice of parameters. The Bayesian prior is defined by the parameters used in the precalculated grid discussion in \S \ref{subsubsec_grid}. The summary statistic that we use is the histogram representations of the data presented in \S \ref{subsubsec_results_DCE_dist} (see Figure \ref{fig_all_DCEs_all_bins}).  We use a simple but strict rejection policy where the forward modeled simulation of the data must match our observed histograms precisely \citep{Marjoram_2015}.

In practice, we search for the
distributions of the input parameters
$\log(\epsilon)$ and $z_{wall}/H$ (see definition in \S \ref{subsubsec_grid})
that 
best
reproduce the observations.
We create a realization of the observed histograms 
by sampling from the cumulative distributions
of the input parameters 
log $\mdot$, cos $i$ and either
log $\epsilon$ or $z_{wall}/H$, depending
on the color excess we aim to reproduce,
and calculating the color excess for
the given combination of model parameters.
We carried out 100
such realizations, each resulting on
a distribution of input parameters
that fitted the observations.
We use the distributions for all 
realizations to calculate a mean
value
and 
standard deviation
at each bin of the
distribution
for each of our parameters of interest.
To calculate the cumulative
distributions of input parameters,
we used the observed mass
accretion rate distributions in each
age bin (\S \ref{subsubsec_observed_mdots})
and
assumed a uniform
distribution of inclinations,
degrees of settling and 
wall heights. 

Depending on the DCE we want to study and reproduce, we use different free parameters. Since the inner wall emission dominates at 4.5 $\mu m$ in the SEDs (cf. Figure \ref{fig_SED}), we use $(z_{wall}/H$, $\log (\dot{M})$, $\cos(i))$  
as the input parameters when we 
attempt to
reproduce the DCEs in the IRAC bands
($J$-[4.5], $J$-[5.8] and $J$-[8.0]). On the other hand, the degree of dust settling is better traced by 24 $\mu m$, where the disk dominates the emission (cf. Figure \ref{fig_SED}), hence, we use $(\log \epsilon$, $\log(\dot{M})$, $\cos (i))$ 
as input parameters when we 
attempt to reproduce the DCEs in the MIPS 24 band
($J$-[24]).

\subsubsection{Results}
\label{subsubsec_results_medians}

\begin{figure}[h!]
\centering
\includegraphics[width=1\linewidth]{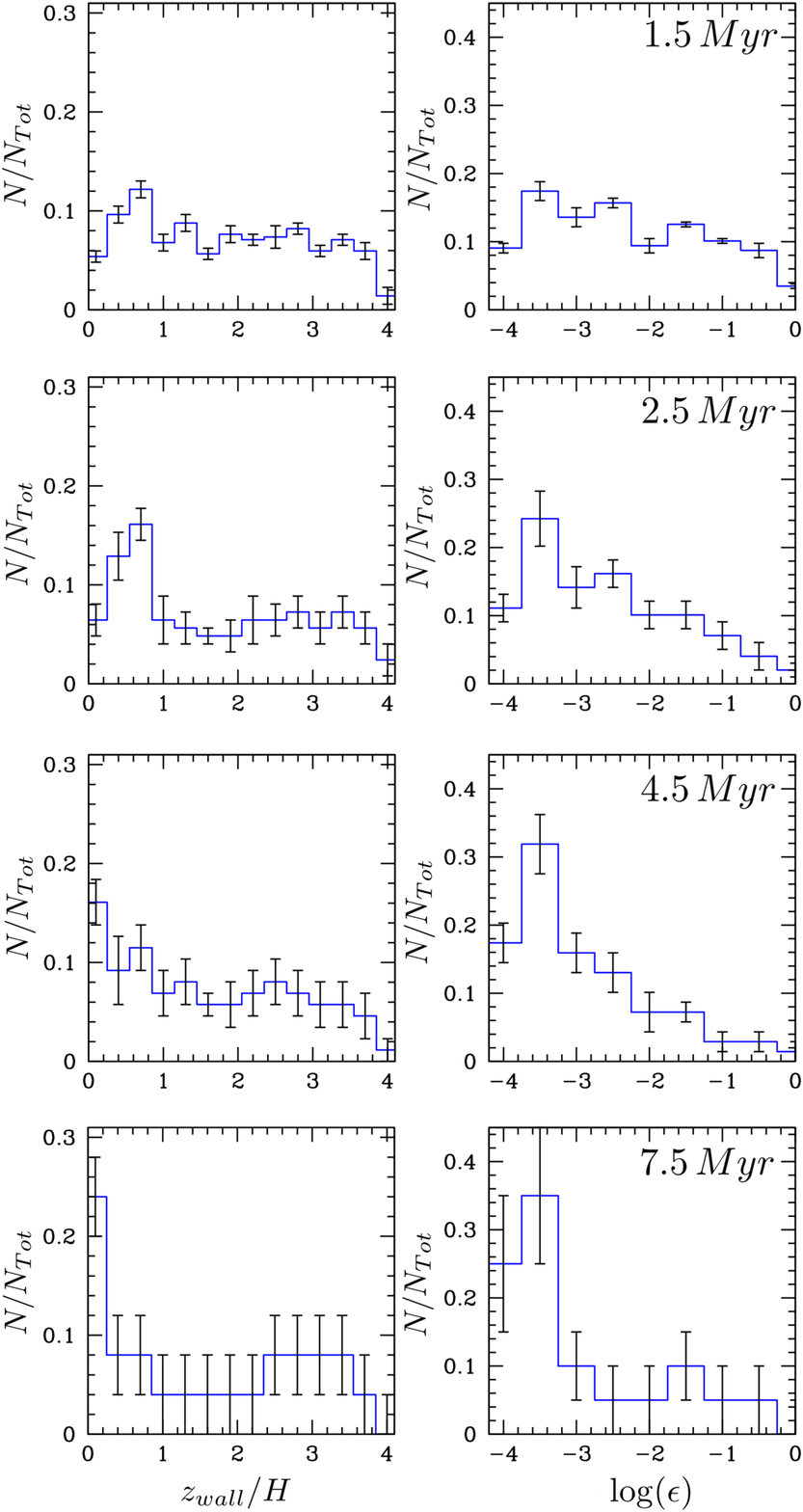}
\caption{\small 
Mean distributions for $z_{wall}/H$ 
(left column) and $\log(\epsilon)$ 
(right column),
for each age bin, with the
youngest bin on the top row.
The distribution of
$z_{wall}/H$ has been calculated
from the color excess $J-[4.5]$ with  $\epsilon = 0.001$ for bins 1, 2, 3 and 4. The distribution of
$\log(\epsilon)$ has been calculated 
from the color excess $J$-[24], with
$z_{wall} = 1.5$ H.
The vertical lines are the standard deviations of each bin.}  
\label{fig_all_parameters_allbins}
\end{figure}

Following the method outlined in
\S \ref{subsubsec_statistical_app},
we inferred the distribution of 
$z_{wall}/H$
from the color excesses $J$-[4.5], $J$-[5.8] and $J$-[8.0],
and the distribution of 
the dust settling parameter
log $(\epsilon)$ from
the $DCE_{J-[24]}$
at each age bin. 
Figure \ref{fig_all_parameters_allbins} shows the 
$z_{wall}/H$ distributions inferred
from $DCE _{J-[4.5]}$ (left column)
and the distributions of
log $(\epsilon)$ (right column)
for each age bin, the youngest located at the top row. We only show the distribution for the $DCE_{J-[4.5]}$ since 4.5 $\mu m$ is the best band to probe the wall height  (see \S \ref{subsubsec_grid} and Figure \ref{fig_SED}).
The distributions for $\log(\epsilon)$ and $z_{wall}/H$ and their peaks move towards lower values as the age of the bin increases (from top to bottom), similarly to what is observed for the mass accretion rates (Figure \ref{fig_observed_mdots_2}) and DCEs (Figure \ref{fig_all_DCEs_all_bins}).

Notice that we use a central star in each age bin, with the corresponding stellar properties ($T_{\text{eff}}, R_{\ast}, L_{\ast}$) associated to the age of the bin, thus the gas scale height $H$ at a fixed radius is different from one bin to the other. The gas scale height is $H\propto M_{\ast}^{-1/2}R^{3/2}$, where $M_{\ast}$ is fixed to 0.30 $M_{\odot}$ (see \S \ref{subsubsec_description_models}) 
for all bins, but the dust destruction radius, $R=R_{in}$, moves inwards with  
age since the luminosity of the star is decreasing. This means that the gas scale height, and thus the wall, is physically
lower
as the age bin increases, since it is measured in units of H, which decreases with age. 
In order to test the robustness of the method, we increased the number of realizations (see \S \ref{subsubsec_statistical_app}) in the analysis 
and no changes in our results were found.
 
 \section{Discussion}
\label{sec_discussion}

\subsection{Implications for disk evolution}
\label{subsec_implications}

\begin{figure}[h!]
\centering
\includegraphics[width=1\linewidth]{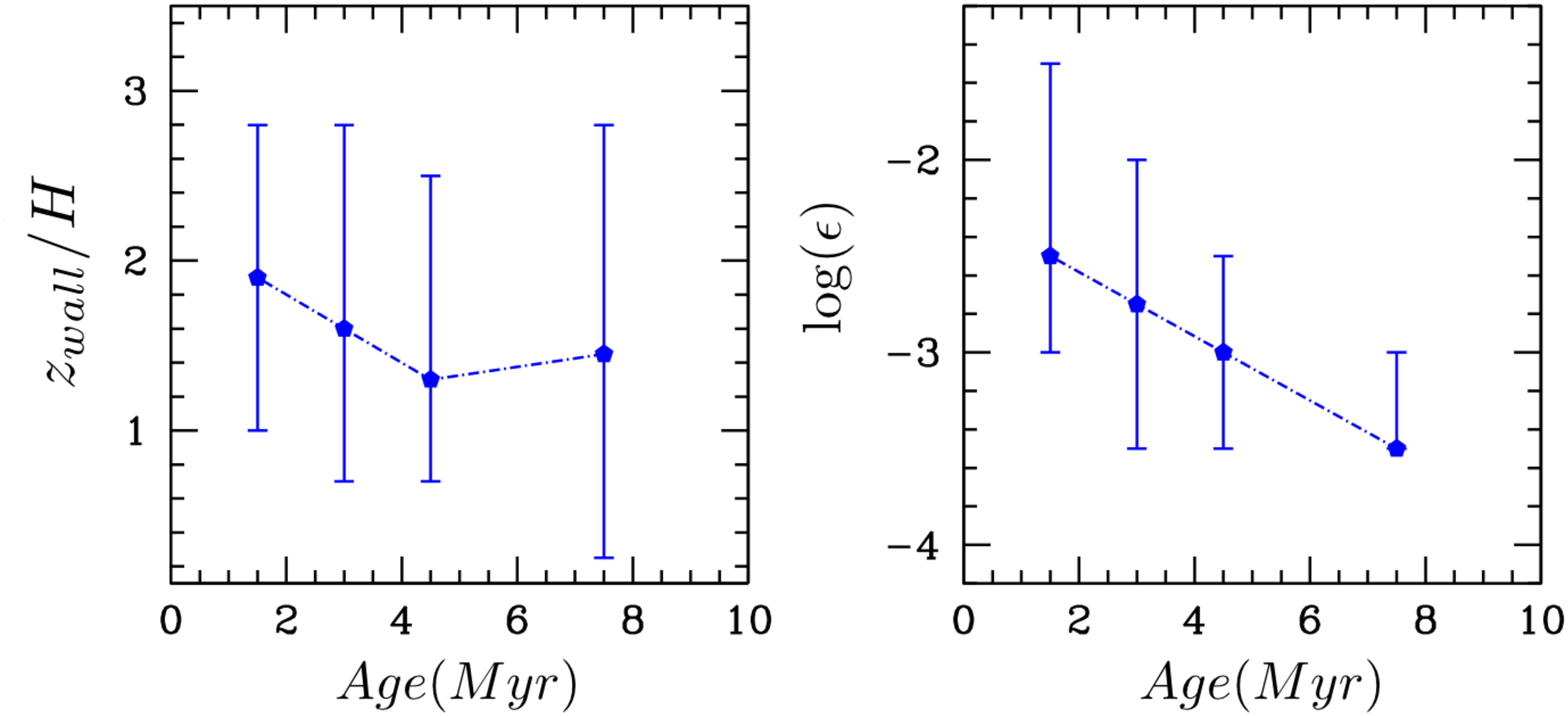}
\caption{\small Median of $z_{wall}/H$ (left panel) obtained from $DCE_{J-[4.5]}$ and median of $log( \epsilon)$ (right panel) obtained from $DCE_{J-[24]}$ vs the age of the bin. The error bars are the quartiles, which show where $50\%$ of the disks lie. In the $z_{wall}/H$ panel, we show the results for a fixed $\epsilon$: 0.001, and in the $\log(\epsilon)$ panel we show the results for a fixed $z_{wall}/H:$ 1.5}  
\label{fig_medianas_zwall_y_eps_vs_age}
\end{figure}

By reproducing 
the
observed distributions of DCEs (\S \ref{subsubsec_results_DCE_dist}) using an extensive grid of disk models (\S \ref{subsubsec_grid}) and taking into account the observed evolution of mass accretion rates (\S \ref{subsubsec_ev_mass_acc_rate}), we have  
inferred
how the 
distributions of the disk parameters $\log(\epsilon)$ and $z_{wall}/H$ (Figure \ref{fig_all_parameters_allbins}) change with age,
comparing results for
the different
age bins.
Figure \ref{fig_medianas_zwall_y_eps_vs_age} shows
 the results of the dust evolution analysis shown
 in \S \ref{subsubsec_results_medians}
 in a different way.
 This Figure 
 shows the median of the parameters $z_{wall}/H$ 
 and
 $\log(\epsilon)$
 vs the age of the bin,
 calculated from the mean and quartiles of the
 mean distributions in
 Figure \ref{fig_all_parameters_allbins}.

\begin{figure}[h!]
\centering
\includegraphics[width=1\linewidth]{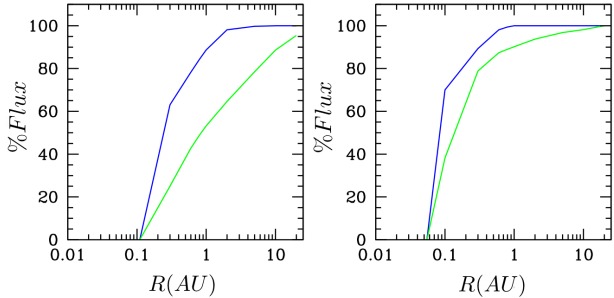}
\caption{\small Emergent flux at 8 $\mu m$ (blue line) and 24 $\mu m$ (green line), at different disk radii, for a disk with $\log(\dot{M})=-8.49 \, M_{\odot}/yr$ with $\epsilon=0.003$ (left panel) and $\log(\dot{M})=-9.29 \, M_{\odot}/yr$ with $\epsilon=0.0003$ (right panel), for a 1.5 (bin 1) and 7.5 Myr old (bin 4) star, respectively.}
\label{fig_fluxes}
\end{figure}

The decrease of
the median of $\log(\epsilon)$ with age in
the right panel in Figure \ref{fig_medianas_zwall_y_eps_vs_age}
indicates
that dust depletion
increases 
with age in the
inner disk regions.
However,
the region probed by the flux
at 24 $\mu$m changes as the
dust gets increasingly depleted
\citep{Dalessio_2006}.
In Figure \ref{fig_fluxes} we show the
contribution to the total flux from the disk region inside radius $R$ as a function of
$R$
at 8 $\mu m$ (blue line) and 24 $\mu m$ (green line)
for disks with the median parameters in age bins
1 and 4.

It can be seen that
$\sim 90\%$ of the flux at 24 $\mu$m arises from the inner $\sim 10$ au of the 
disk at age bin 1 with
$\epsilon=0.003$, while
the same flux fraction arises
from the  inner $\sim 1$ au of the disk for 
age bin 4, with $\epsilon = 0.0003$.
In both cases, we are probing
 the inner few au of the
disk ($\sim 1-10$ au), which is the region where terrestrial planets form.

The low values of
$\log(\epsilon)$ in Figure \ref{fig_all_parameters_allbins}, (right panels)
indicate that 
even for the youngest groups, disks 
already show 
significant  
dust settling, 
suggesting 
that the settling process starts
at very early evolutionary stages ($\sim 1$ Myr).

Our results 
support the
conclusions
derived from
analysis of IRS spectra in
Taurus \citep{Furlan_2006,Furlan_2009},
Ophiuchus \citep{McClure_2010}, and
Chamaeleon \citep{Manoj_2011}. \citet{Furlan_2009} find values of $\log(\epsilon)\sim -2$ to -3 in Taurus, Chamaeleon I \citep[2 Myr;][]{Luhman_2004} and Ophiuchus \citep[$<1$Myr;][]{Luhman_1999}. These values are consistent with the median of $\log(\epsilon)$ shown in Figure \ref{fig_medianas_zwall_y_eps_vs_age} for bin 1 and 2, indicating a large degree of settling in disks.
Similarly, our results
are consistent with the
results of the
analysis of Herschel/PACS
data of disks
in the (1.5 Myr) Lynds 1641 region by
\citet{Grant_2018} and in the (3 Myr) $\sigma$ Ori cluster by \citet{Mauco_2016},
who found 
a significant fraction of disks with $-2.8 \lesssim \log (\epsilon) \lesssim -1.8$.

The other important parameter we have considered is
$z_{wall}/H$, 
which is related to the height of 
the disk wall $Z_{wall}$ or
dust sublimation front, located at the radius where the disk reaches the 
temperature above which silicates sublimate
(\S \ref{subsubsec_grid}.)
We have used the excess $DCE_{J-[4.5]}$ as a probe of the wall height. This is the best band for this purpose, because
as can be seen in Figure \ref{fig_SED} where the flux at 4.5 $\mu$m is indicated,
 the flux at longer wavelengths (5.8 $\mu m$ or 8.0 $\mu m$) 
 includes 
 contributions from the disk, while 
 at 
 3.6 $\mu m$ 
 the color 
 excess might be more contaminated by the star.

\begin{deluxetable*}{ccccc}
	\tablecaption{Range of values of $Z_{wall}$, and mode of the distribution of $z_{wall/H}$ in each age bin
	\label{tab_zwall}}
\tabletypesize{\scriptsize}
\tablehead{
	\colhead{Age} &  
	\colhead{mode($z_{wall}/H$)}  &
	\colhead{$Z_{wall_{min}}$} & 
	\colhead{$Z_{wall_{max}}$}  &
	\colhead{Age}\\
	\colhead{Myr} &  
	\colhead{}  &
	\colhead{au} & 
	\colhead{au}  &
	\colhead{bin} 
}
\startdata
 1.5 & 0.6 & $3\times 10^{-3}$  & $5.85\times 10^{-3}$ & 1\\ 
 2.5 & 0.6 & $2.02\times 10^{-3}$ & $5.85\times 10^{-3}$ & 2\\ 
 4.5 & 0.2 & $5.1\times 10^{-4}$ & $2\times 10^{-3}$ & 3\\ 
 7.5 & 0.2 & $4.22\times 10^{-4}$ & $2\times 10^{-3}$ & 4\\
\enddata
\end{deluxetable*}

In Figure \ref{fig_medianas_zwall_y_eps_vs_age} the median of $z_{wall}/H$ is $\sim 2, 1.6, 1.3$ and 1.45 at 1.5, 2.5, 4.5 and 7.5 Myr, respectively; thus it decreases from bin 1 to bin 3 without a significant change at bin 4. Although the median of $z_{wall}/H$ is taken as a representative value of the corresponding distributions,
Figure \ref{fig_all_parameters_allbins}
shows that at all age bins there is a wide range of values of $z_{wall}/H$. The peak of the distribution for $z_{wall}/H$ in Figure \ref{fig_all_parameters_allbins} 
is located at $ z_{wall}/H \sim 0.6$, 0.6, 0.2 and 0.2 for age bins 1, 2, 3 and 4, respectively, and it becomes more pronounced as the age increases. In Table \ref{tab_zwall} we show the mode of the distributions of $z_{wall}/H$ shown in Figure \ref{fig_all_parameters_allbins}, and the corresponding range of values of
$Z_{wall}$ in each age bin, given by $Z_{wall_{min}}$ and $Z_{wall_{max}}$.
As can be seen,
even if $z_{wall}/H$ stays relatively constant in the older age bins, the physical height of the wall decreases
with age (see Table \ref{tab_zwall}). This is because the scale height $H \propto R_{in}^{3/2}$, the sublimation radius
$R_{in} \propto (L_* + L_{\text{acc}})^{1/2}$, 
and the stellar and accretion luminosity
decrease with age. The height of the wall is also an indicator of 
dust settling in the innermost regions of the disk, so our results
indicate that
the dust in these regions also 
settle with age.

The criteria we applied in this work to select the observational sample
allows us to eliminate PTDs and TDs with large cavities devoid of optically thin dust, i.e., disks with only photospheric near-IR emission.
However,
we may have included
in our analysis 
a few transitional disks with inner cavities filled with 
sufficient
optical thin
material 
to produce an
excess at 5.8 $\mu m$ and 8 $\mu m$ 
that satisfies our observational selection criteria (see \S \ref{subsec_DCE}).
This is particularly
relevant for the older bins, but according to \citet{Espaillat_2014}, TDs represent $\sim 10-20 \%$ of the total population of disks in different stellar groups an this fraction reduces when considering only transitional disks with optical thin
material inside their cavities.
Moreover disks with inner clearings
have relatively
high fluxes at 
24 $\mu$m arising in the
frontally illuminated
edge of the cavity
\citep{Grant_2018};
in contrast,
our statistical analysis
indicates 
that
the flux at
24 $\mu$m  is sufficiently low to be consistent with a very high degree of settling. A scenario in which the disk flux decreases at all bands due to increased dust settling and growth is more consistent with our evolutionary results. 

Although we cannot determine the conditions under which the disks follow the evolved disks scenario vs inner disk clearing, we find that from the initially compiled sample containing diskless as well as disk-bearing stars in all the stellar groups, before applying the selection criteria described in \S \ref{subsec_DCE} to define our disk-bearing stars sample, $\sim 35 \%$ of them were classified as full- disk-bearing stars which have undergone the evolved disk path thus, according to their SEDs, they do not have large inner cavities as transitional disks. The other $\sim 65 \%$ of the stars are either diskless stars, TDs or objects at earlier evolutionary stages such as Class I objects. We examined the sample of stars rejected by our selection criteria from the initial sample, searching for transitional disks -
that is, 
disks without near-IR excess but with 24 $\mu m$ comparable to that of the Taurus median \citep{Mauco_2016} - and found 4, 4, 5 and 2 TDs, in the age bins 1, 2, 3, and 4, respectively.
These transitional disks represent $\lesssim 1\%$ of the initially compiled sample,
in agreement with the low occurrence found in other studies
\citep{Espaillat_2014}.

\subsection{Disk frequency}

Observations of 
disk frequency \citep{Hernandez_2007a, Briceno_2019},  
indicate that $\sim 70-80\%$ of the stars in young ($\sim 1$ Myr) stellar groups are disk-bearing stars, while at 5 Myr this fraction drops to $\sim 15\%$. For groups older than $\sim 8$ Myr, only about $\sim 8\%$ of the disk-bearing stars have survived.  

\begin{figure}[h!]
\centering
\includegraphics[width=1\linewidth]{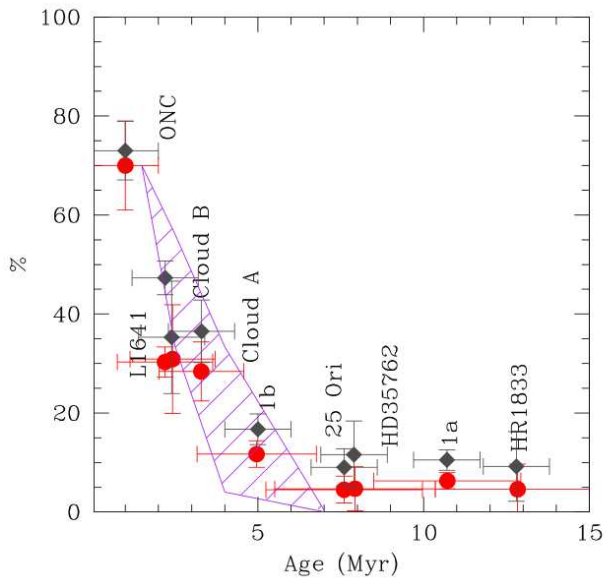}
\caption{\small 
Predicted disk frequency as a function of age for disks following viscous evolution and photoevaporation.
The purple region shows predictions for
$\dot{M}_{\text{phot,lim}} \sim 3 \times 10 ^{-9} \, M_{\odot}/yr$ (lower boundary) and
$ \dot{M}_{\text{phot,lim}} \sim  10 ^{-9} \, M_{\odot}/yr$ (upper boundary).
The fraction of accretors (red dots) and the disk fraction (dark gray diamonds) as a function of age for different stellar groups 
taken from \citet{Briceno_2019}
are shown for comparison. 
}
\label{fig_frac_discos}
\end{figure}

In Figure \ref{fig_mdot_model} we show that the observed decay of the mass accretion rate is consistent with viscously evolving disks; however, viscous evolution 
cannot 
explain the decrease of disk
frequency, since it only predicts that the disk will expand while decreasing its surface density with age, and a fraction of the disk material will be accreted onto the star. Another agent driving disk evolution
is photoevaporation. 
The upper levels of the disk are heated by high
energy radiation from the star and accretion shock
so that the sound speed becomes higher
than the escape velocity. The gas leaves the disk
in a photoevaporative wind, at a rate that depends
on the radiation field 
considered to be more effective in driving the wind,
either X-rays, FUV or EUV radiation
\citep[][and references therein]{Alexander_2014}.
 Since the mass accretion
 rate decreases with time
 as a consequence of disk evolution, 
 it eventually becomes comparable to the mass loss rate in the wind; when this happens, matter from the outer disk flows away instead of reaching the star, while the 
 inner disk quickly drains out onto the star 
 \citep{Clarke_2001}.
 Models of photoevaporative winds predict after this
 time is reached, the rest of the disk dissipates in time scales smaller than the viscous time scale \citep{Alexander_2007,Alexander_2014}.

We have attempted to 
incorporate the effects of photoevaporation
in our 
study of disk evolution in a simple
way. Essentially, 
we allow the disks around the stars in the first
age bin to evolve viscously, and assume that they dissipate instantaneously once their mass accretion rate reaches a given limiting value, which would correspond to the
photoevaporation
mass loss rate.
To do this we take the observed distribution of mass accretion rates at bin 1 (upper left panel in Figure \ref{fig_observed_mdots_2}) and assume a disk fraction of $70 \%$ at bin 1 (1.5 Myr) \citep{Briceno_2019}. We set a limit for the photoevaporation rate $\dot{M}_{\text{phot,lim}}$ such that if $\dot{M}<\dot{M}_{\text{phot,lim}}$, 
the disk stops 
accreting and essentially disappears.
We let each bin from the initial distribution of $\dot{M}$ viscously evolve following equation  (\ref{eq_mdot0}) with $\alpha$=0.01, $R_{1}$=10 au and $M_{d}(0)$=0.07 $M_{\odot}$ for a $0.30 \, M_{\odot}$ star, which are the values used in constructing the green solid line in Figure
\ref{fig_mdot_model}, but we note that any value of the triad ($M_{d}(0)$, $R_1$ and $\alpha$) that fits the observed $\mdot({\rm age})$ with similar $\chi^2$ (see Fig. \ref{fig:chisquared}) leads to the same result, by construction.
At each age bin we eliminate the disks with mass accretion rates below $\dot{M}_{\text{phot,lim}}$ and recalculate the disk fraction. 
We vary $\dot{M}_{\text{phot,lim}}$ 
to achieve a good fit to
the observed disk fractions.
We find that values of
$\dot{M}_{\text{phot,lim}}$
between 
$\sim 10 ^{-9} M_{\odot}/yr$
and
$\sim 5 \times 10 ^{-9} M_{\odot}/yr$
give a reasonable
fit to the observations,
as shown in Figure \ref{fig_frac_discos},
in which we plot the disk fraction 
and the accretors fraction  
for different stellar groups as a function of age
from \citet{Briceno_2019}. A higher value of $\dot{M}_{\text{phot,lim}}$ would result in a more rapid decay of the disk frequency
because disks would disappear much sooner.
 
As shown in
Figure \ref{fig_frac_discos}
the predicted disk fraction as a function of age 
follows the observations reasonably well. 
Our simple treatment predicts no inner disks beyond 7 Myr, possibly because we did not
include other mechanisms that may allow the existence of inner disk at these older ages. Disks surviving at ages $> 7$ Myr may be those that started with higher masses, but detailed calculations including the actual evolution of stars in the presence of photoevaporation are required for a better comparison with observations. 
A more 
detailed estimate of the disk frequency will be made in a future work in order to obtain a disk fraction more consistent with the observed values, in particular in the oldest groups.
In any case,
our results together with Figure
\ref{fig_mdot_model}, 
suggest that 
the main agents driving
evolution of the
gas in the disk
are
viscous evolution 
and photoevaporation.

\subsection{Snowlines}

The gas component of the disk contains atoms and molecules of different species, some of which sublimate at relatively low temperatures ($\lesssim 150$ K); these species are known as {\it volatiles}. These volatiles seem to play a major role in the process of planet formation, giving rise to an active chemistry that influences the composition of the planets' atmospheres during their formation process. The disk radius where a certain volatile reaches its sublimation temperature is called snowline or condensation front. Outside the snowline the volatiles are frozen, adhered to the surface of dust grains, and inside the snowline they remain in the gas phase. In the last decade, studying the location of different snowlines has gained attention, in particular to explain the chemical composition of some exoplanetary atmospheres, since it depends directly on the available molecules in the surrounding regions of the planets, during their formation process \citep{Oberg_2011, AliDib_2014, Madhusudhan_2014}. It has also been suggested that an enhancement of dust growth can occur near the location of different condensation fronts, favoring the accumulation of mm to dm grains, which might eventually form planetesimals \citep{Zhang_2015, Okuzumi_2016}.

\begin{figure*}[h!]
\centering
\includegraphics[width=1\linewidth]{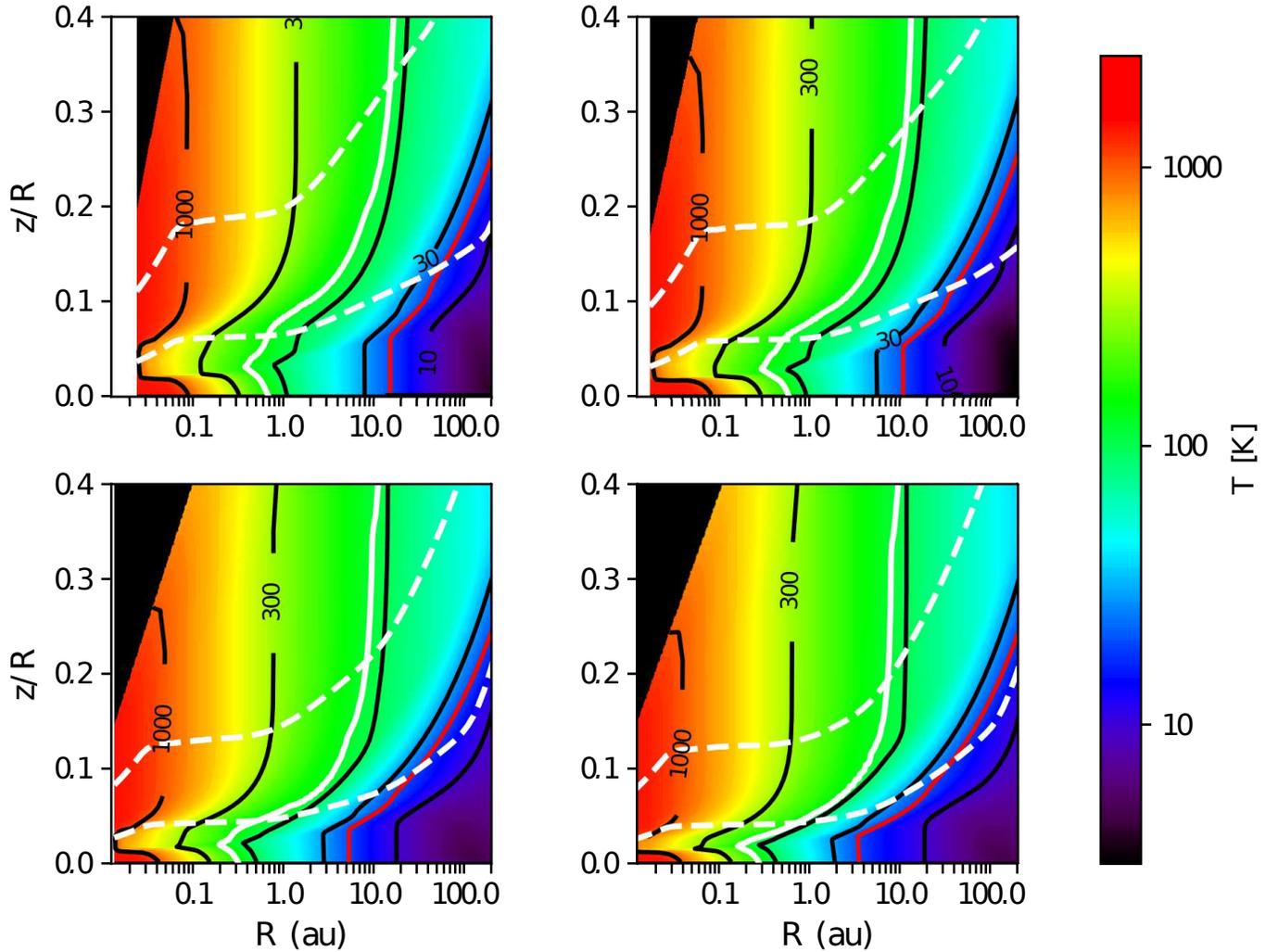}
\caption{\small Location and shape of the \H2O (white solid line) and CO (red solid line) snowlines for 
disk models calculated for the median mass accretion rate $\dot{M}$ and median settling parameter $\epsilon$ at
each age bin, 
with bin 1 (upper left), bin 2 (upper right), bin 3 (lower left) and bin 4 (lower right). The color scale shows the temperature distribution with temperature iso-contours for 10, 30, 100, 300, 1000 K (black lines). The white dashed lines show the location of 1 and of 3 scale heights at each radius. 
}
\label{fig_snowlines}
\end{figure*}

Some authors have found that the shape and location of the different snowlines are influenced by changes in the gas surface density \citep{Piso_2015, Krijt_2016, Powell_2017}. Molecules such as CO and its isotopologues have been observed and measured \citep{Qi_2011, Qi_2013, Qi_2015}, and while the \H2O snowline is far more difficult to detect and study because it is located very close to the star, they are both important to understand their influence in the dust component of the disk, the chemical abundances in exoplanetary atmospheres, and also the composition of the planets and asteroids in the Solar System.  A more general overview about the different theoretical and observational aspects of snowlines can be found in \citet{Pontoppidan_2014}.

We study the location and evolution of different snowlines, in particular \H2O and CO, under the assumption that the gas and the dust in the disk follow the same temperature distribution. 

\begin{figure}[h!]
\centering
\includegraphics[width=0.9\linewidth]{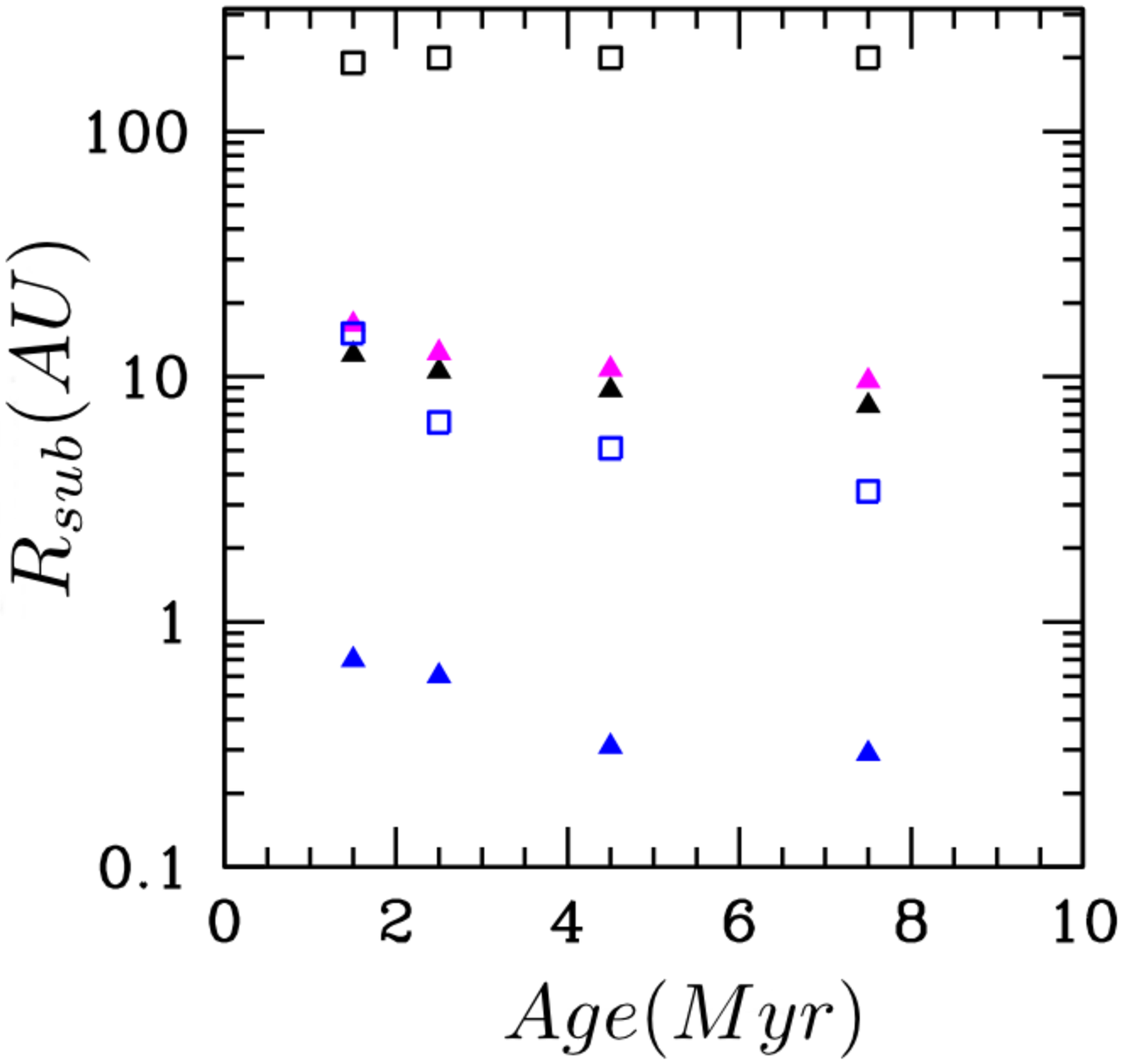}
\caption{\small Location of the \H2O (solid triangles) and CO (empty squares) snowlines, for different values  of $z/R:$ 0 (blue), 0.25 (black) and 0.4 (magenta).}
\label{fig_R_vs_age_2}
\end{figure}

For each model in our grid we calculate self consistently the vertical and radial structure of the disk (see \S \ref{subsubsec_dalesio_model}),
which allows us to find the shape and location of the different snowlines. We take the density-dependent sublimation temperature for water ice reported in Table 3 from \citet{Pollack_1994} and for CO 
we adopt 20 K 
following \citet{Oberg_2011} and \citet{Qi_2011, Qi_2013, Qi_2015}. Figure \ref{fig_snowlines} shows the shape and location of the \H2O and CO snowlines, for disk models, which we refer to as mean models, calculated using the median observed values of the mass accretion rate (from Table \ref{tab_mdot} and Figure \ref{fig_median_logmdot_vs_age}) and the median $\epsilon$ parameter found in this work (right panel in Figure \ref{fig_medianas_zwall_y_eps_vs_age}) at the different age bins. 
The snowlines are not vertical sublimation fronts,
but they curve 
towards large radii as the
height above 
the midplane increases, which is a consequence of the vertical temperature gradient in irradiated disks in which the temperature increases towards the disk surface \citep{Dalessio_1998, Dalessio_2006}.
As disks become older, the midplane snowline moves inwards because the stellar and the accretion luminosity decrease (Fig. \ref{fig_median_logLum_vs_age_susana}) and the inner disk becomes colder, allowing certain volatiles to remain frozen closer to the star. There is an
 additional effect in the \H2O snowline which is more accentuated in the disk models with the highest mass accretion rates (e.g. at bin 1), in which the snowline moves inwards as $z/R$ decreases from the disk surface to the midplane, but at $0 \lesssim z/R \lesssim 0.03$ is pushed outwards, from $
 \sim$ 0.4 au at $z/R \sim 0.03$ to $\sim$ 0.7 au at the very midplane $(z/R=0$). At bin 4 the midplane snowline is pushed from $\sim$ 0.17 au at $z/R \sim 0.02$ up to 0.29 au at $z/R=0$. This effect is stronger in the first two age bins and it is a consequence of  viscous dissipation, since the higher the
  mass accretion rate, the more efficient the viscous heating in this region \citep{Dalessio_1998,Dalessio_2001, Dalessio_2006, Drazkowska_2017, Huang_2017, Pinte_2018}. Figure \ref{fig_snowlines} also shows the location of 1 and of 3 scale heights $H$ (white dashed lines), evaluated at the midplane temperature; it can be seen that the scale height decreases with age. Figure \ref{fig_R_vs_age_2} shows the location of the \H2O and CO
   snowlines; this value depends on $z/R$ as already noticed, thus we show the location for different values of $z/R$: 0 (blue), 0.25 (black), and 0.4 (magenta). The snowline location decreases with age, except for the case $z/R=0.25$ for the CO snowline, where it remains approximately constant.

\begin{deluxetable}{lcc}
	\tablecaption{Midplane location of the \H2O and CO snowlines
	\label{tab_snowlines}}
\tabletypesize{\scriptsize}
\tablehead{
	\colhead{Age} & 
	\colhead{$R_{\text{sub}, \text{H}_2\text{O}}$} &  
	\colhead{$R_{\text{sub}, \text{CO}}$}\\ 
	\colhead{Myr} & 
	\colhead{au} &  
	\colhead{au} 
}
\startdata
1.5 & 0.7 & 15\\ 
2.5 & 0.6 & 7\\ 
4.5 & 0.31 & 5\\
7.5 & 0.29 & 3.5\\
\enddata
\end{deluxetable}

 Table \ref{tab_snowlines} shows the midplane snowline location for \H2O and CO at the different age bins. At 1.5, 2.5, 4.5, and 7.5 Myr the water snowline is located at $\sim$ 0.7, 0.6, 0.31, and 0.29 au, respectively, while the CO snowline is located at $\sim$ 15, 7, 5, and 3.5 au.
 \citet{Qi_2013} reported a $R_{\text{sub},\text{CO}}\sim 30$ au for the $\sim 10$ Myr old disk around TW Hya, whereas \citet{Schwarz_2016} located the CO midplane snowline at $\sim 17-23$ au, using ALMA observations of CO isotopologues, for the same source. Notice that the disk around TW Hya is a transitional disk, with an inner hole of $\sim 4$ au \citep[see][]{Espaillat_2014}, which should be hotter than the disks from our mean models, since TW Hya has a 0.8 $M_{\odot}$ star which irradiates directly the disk wall, while our calculations were made for a 0.3 $M_{\odot}$ star surrounded by a full disk; and this might explain the difference in the locations we find for the CO snowline. The location of the snowlines, determined with observations, is sensitive to the disk depth at which the emission of the volatiles is being detected; our models show that at $z/R \sim 0.06$ au, $R_{\text{sub}, \text{CO}}\sim 13$ au while at $z/R \sim 0.12$ au, $R_{\text{sub}, \text{CO}}\sim 40$ au at bin 4 (7.5 Myr). \citet{Mulders_2015} calculate the \H2O snowline  for disks with different dust grain sizes, and find an analytical expression for two regimes of the maximun grain sizes: ISM-like and cm-sized grains. The values they find range from 0.28-0.60 au (1.5Myr), 0.22-0.49 au (2.5 Myr), 0.11-0.23 au (4.5 My) and 0.07-0.16 au (7.5 Myr) for a $0.30 \, M_{\odot}$ star, which are similar to our results
(Table \ref{tab_snowlines}). The location of the snowlines at the different age bins may put constrains on the timescale for the formation of planets within the inner au in the disk.

\section{Summary and conclusions}
\label{sec_summary_conclusions}

\begin{itemize}
\item  We performed a statistical study focused on the global evolution of protoplanetary disks around T Tauri stars, that includes observed mass accretion rates as a constrain to infer the inner disk properties. The observational sample includes various stellar groups with IRAC and MIPS 24 photometry, as well as PANSTARRS $[g-i]$ color (used to calculate \Av uniformly for most stars),
with ages from 1 to 11 Myr, and located at distances $\lesssim 500$ pc.

\item By introducing the disk color excess (DCE) as an indicator of infrared excess, we study the evolution of the observed IR emission from the disk-bearing stars in the stellar groups, separated into different age bins (1.5, 2.5, 4.5 and 7.5 Myr) to make a more robust statistical study at each age. The median of the different DCEs decreases as the age of the bin increases.

\item  We use a different sample of TTSs with significant overlap with the DCEs sample, and with additional GAIA DR2 distances, to determine mass accretion rates using the EW of the H$\alpha$ line, which was either available from previous works, or determined by us. The observed decay of $\dot{M}$ with age is consistent with viscously evolving disks; in particular we find a good fit to the observed data for a viscosity coefficient $\alpha=$0.01, an initial disk mass $M_{d}(0)=0.07 \, M_{\odot}$, and a characteristic radius $R_{1}=$10 au, inside which 60\% of the initial disk mass is located. Additionally,  different combinations of the triad ($M_{d}(0)$, $R_{1}$, $\alpha$) can also provide a good match as determined from the $\chi^{2}$ value in Figure \ref{fig:chisquared}.

\item Using the D'Alessio irradiated accretion disk (DIAD) models, we computed, for each age bin, a grid of disk models varying the degree of dust settling $\epsilon$, the mass accretion rate $\dot{M}$, the parameter $z_{wall}/H$ related to the height of the wall, and the cosine of the inclination angle $\cos(i)$. We incorporated a two-layered wall to emulate the wall curvature, as well as an exponential tapering at the outer edge of the disk. Using this grid we computed synthetic DCEs in order to forward model the observed DCEs distributions. The observed $\dot{M}$ distributions at the different age bins are implemented as constrains in the method.

\item The distributions of the parameters $\epsilon$, which measures the dust depletion in the upper disk layers,  and $z_{wall}/H$, which measures the height of the wall at the dust destruction radius, show that even at 1.5 Myr (bin 1) a fraction of the disks are highly settled ($\log(\epsilon) \lesssim $ -3.0) with low walls ($z_{wall}/H\lesssim 1$), and the fraction of these disks increases with age; also the peak of the distributions moves towards smaller values.

\item We make a first attempt to test photoevaporation by letting the observed distribution of $\dot{M}$ from bin 1 viscously evolve, and we set a photoevaporation mass loss rate $\dot{M}_{\text{phot,lim}}$ below which the star stops accreting material from the disk. We find that $\dot{M}_{\text{phot,lim}} \sim 1 - 3 \times 10^{-9} \, M_{\odot}yr^{-1}$ fits reasonably well the observed decay of the disk and accretors fractions with age.

\item We find that the \H2O and CO snowlines curve towards 
larger radii as the height over the disk midplane increases
because of the vertical temperature gradient. At $0 \lesssim z/R \lesssim 0.02$, the \H2O snowline is pushed outwards due
to viscous dissipation.
The snowlines migrate inwards as the disk evolves, as a consequence of the decrease in the stellar and accretion shocks luminosity with age. The \H2O midplane snowline starts at $\sim$ 0.7 au at 1.5 Myr and ends up at $\sim$ 0.29 au at 7.5 Myr; the CO midplane snowline starts at $\sim$ 15 au at 1.5 Myr and ends up at $\sim$ 3.5 au at 7.5 Myr for the 
the models with the median mass accretion rate $\dot{M}$ and median degree of settling $\log(\epsilon)$ at each age bin.
\end{itemize}

We thank an anonymus referee for comments and suggestions that greatly improved the presentation of this paper. We also thank Tom Megeath for providing infrared photometry for the ONC, Kevin Luhman for providing intrinsic photospheric colors for early K-type stars, and Lee Hartmann and Jaehan Bae for insightful conversations. E.M.M. and S.L. acknowledge support from PAPIIT-UNAM IN101418 and CONACyT 23861. E.M. M. acknowledges support from a CONACyT scholarship. J.H  acknowledges support from PAPIIT-UNAM IA102319. N. C. acknowledges partial support from NASA grant NNX17AE57G.
K.M. acknowledges financial support from FONDECYT-CONICYT project no. 3190859 and from the ICM (Iniciativa Cient\'ifica Milenio) via the N\'ucleo Milenio de Formaci\'on Planetaria grant.

\startlongtable
\movetabledown=0.3in
\begin{longrotatetable}
\begin{deluxetable*}{ccccccccccccccc}
\tabletypesize{\scriptsize}
\tablewidth{0pt}
\tablecaption{Stellar properties \label{megatable}}
\tablehead{
\colhead{2MASS} & 
\colhead{SpT} & 
\colhead{A$_V$} & 
\colhead{$T_{\rm eff}$(K)}  &
 \colhead{Ref} &
  \colhead{$L_{\ast}/L_{\odot}$} &
 \colhead{$M_{\ast}/M_{\odot}$} &
  \colhead{$R{\ast}/R_{\odot}$} &
  \colhead{dist(pc)}  &
  \colhead{EW {H$\alpha$}(\AA)} &
  \colhead{$DCE_{J-[4.5]}$} &
  \colhead{$DCE_{J-[5.8]}$} &
 \colhead{$DCE_{J-[8.0]}$} & 
 \colhead{$DCE_{J-[24]}$} & 
 \colhead{$\dot{M}(M_{\odot}yr^{-1})$} 
}
\startdata
\cutinhead{ONC}
 05341189-0506162   &    M5.5      &      0.55   &       2740    &       1   &       0.03    &       \nodata   &       \nodata   &      416     &       -105.8   &    0.49     $ \pm $ 0.04         &      0.88     $\pm$ 0.05         &     1.69   $\pm$ 0.05      &       4.2       $\pm$ 0.09           &         \nodata                      \\ 
 05342125-0450326   &    M3.5      &      1.4    &       3260    &       1   &       0.03    &       0.18      &       0.58      &      245     &       -15.0    &    0.21     $\pm$ 0.03          &     0.55     $\pm$ 0.04          &      1.61     $\pm$ 0.04       &          6.16      $\pm$ 0.04           &    $    1.06\times10^{-10}   $       \\
 05372601-0534013   &    K3.5      &      0.6    &       4440    &       2   &       0.81    &       1.14      &       1.52      &      382     &       -3.0     &      \nodata                     &       \nodata                     &       \nodata                  &           \nodata                       &    $    2.91\times10^{-10}   $         \\
05334954-0536208   &    K5.5      &      1.0    &       4080    &       1   &       0.43    &       0.81      &       1.31      &      404     &       -5.8     &     1.73     $\pm$ 0.02          &      2.19     $\pm$ 0.02          &      2.87     $\pm$ 0.02     &          4.77      $\pm$ 0.03          &    $    3.30\times10^{-10}   $       \\
05352720-0530247   &    M5.0      &      1.69   &       2880    &       4   &       0.44    &       \nodata   &       \nodata   &      415     &       \nodata  &    0.23    $ \pm$ 0.05          &      0.58     $\pm$ 0.05          &     1.43     $\pm $0.05       &           \nodata                       &         \nodata                    \\    
\cutinhead{Taurus}
04270469+2606163  &   K6.0  &    0.0 &   4020   &  5   &     \nodata &     0.72    &     1.92    &       121     &     -42.8   &     \nodata                      &     \nodata                      &     \nodata                       &     \nodata                       &     $ 8.02\times10^{-9}  $       \\
04141700+2810578  &   K3.0  &    2.36 &   4550   &  5   &     1.03    &     1.21    &     1.63    &       132     &     -90.7   &      2.98   $ \pm$ 0.07          &      3.32    $\pm$ 0.07          &       3.88    $\pm$ 0.07          &       6.61    $\pm$ 0.08          &     $ 1.15\times10^{-8}  $       \\
04231822+2641156  &   M3.5  &    4.4  &   3260   &  5   &     \nodata &     \nodata &     \nodata &       \nodata &     \nodata &      1.24    $\pm$ 0.1          &      1.55    $\pm$ 0.1           &       2.22    $\pm$ 0.1           &       5.5    $\pm$ 0.11          &       \nodata                    \\
\enddata
\tablecomments{This table is published in its entirety in the electronic edition of the {\it Astrophysical Journal}. 
A portion is shown here for guidance regarding its form and content.}
\tablecomments{The errors associated to the DCEs are based only on the photometric errors and thus they are underestimated.}
\tablecomments{ References for spectral types (column 5): 1. Hern\'andez et al. 2020a (in preparation), 2. \citet{Briceno_2019}, 3. \citet{Hillenbrand_1997}, 4. \citet{Hillenbrand_2013}, 5. \citet{Esplin_2014}, 6. \citet{Lada_2006}, 7. \citet{Hernandez_2014}, 8. \citet{Bayo_2008}, 9. \citet{Bayo_2011}, 10. \citet{Luhman_2012}, 11. \citet{Spina_2014}, 12. This work. }
\tablecomments{ The data for the EW {H$\alpha$} comes from: Hern\'andez et al. 2020a (in preparation) (ONC), this work (Taurus), \citet{Hernandez_2014} ($\sigma$ Ori) and \citet{Briceno_2019} (Ori OB1b and Ori OB1a).}
\end{deluxetable*}
\end{longrotatetable}

\bibliographystyle{aasjournal}
\bibliography{EM20}

\end{document}